# Decoding Drug Discovery: Exploring A-to-Z *In silico* Methods for Beginners

Hezha O. Rasul[1 (Corresponding Author)], Dlzar D. Ghafour[2,3], Bakhtyar K. Aziz[4], Bryar A. Hassan[5,6], Tarik A. Rashid[5], and Arif Kivrak[7]

[1]Department of Pharmaceutical Chemistry, College of Science, Charmo University, Peshawa Street, Chamchamal, 46023, Sulaimani, Iraq

[2]Department of Medical Laboratory Science, College of Science, Komar University of Science and Technology, 46001, Sulaimani, Iraq

[3]Department of Chemistry, College of Science, University of Sulaimani, 46001, Sulaimani, Iraq

[4]Department of Nanoscience and Applied Chemistry, College of Science, Charmo University, Peshawa Street, Chamchamal, 46023, Sulaimani, Iraq

[5]Computer Science and Engineering Department, School of Science and Engineering, University of Kurdistan Hewler, KRI, Iraq

[6] Department of Computer Science, College of Science, Charmo University, Peshawa Street, Chamchamal, 46023, Sulaimani, Iraq

[7]Department of Chemistry, Faculty of Sciences and Arts, Eskisehir Osmangazi University, Eskişehir, 26040, Turkey

Email (corresponding): hezha.rasul@chu.edu.iq

## Abstract

The drug development process is a critical challenge in the pharmaceutical industry due to its time-consuming nature and the need to discover new drug potentials to address various ailments. The initial step in drug development, drug target identification, often consumes considerable time. While valid, traditional methods such as *in vivo* and *in vitro* approaches are limited in their ability to analyze vast amounts of data efficiently, leading to wasteful outcomes.

To expedite and streamline drug development, an increasing reliance on computer-aided drug design (CADD) approaches has merged. These sophisticated *in silico* methods offer a promising avenue for efficiently identifying viable drug candidates, thus providing pharmaceutical firms with significant opportunities to uncover new prospective drug targets.

The main goal of this work is to review *in silico* methods used in the drug development process with a focus on identifying therapeutic targets linked to specific diseases at the genetic or protein level. This article thoroughly discusses A-to-Z *in silico* techniques, which are essential for identifying the targets of bioactive compounds and their potential therapeutic effects. This review intends to improve drug discovery processes by illuminating the state of these cutting-edge approaches, thereby maximizing the effectiveness and duration of clinical trials for novel drug target investigation.

# 1. Introduction

The search for new chemical entities with therapeutic potential is known as drug discovery. One of the main objectives of drug development programs is the identification of novel molecular entities that may be effective in treating diseases with specific medical needs. Currently, the medications on the market represent a relatively modest number of pharmacological target categories [1].

There are many moving parts in the drug development process. At its most fundamental level, the process can be broken down into two distinct phases, as shown in Figure 1. The first, known as drug discovery, is conducting experiments and investigations to find a single chemical that can be used in clinical practice after identifying a biological target and related disease state. Target identification, lead discovery, and optimization are three distinct phases that follow the drug discovery phase [2]. Drug discovery relies on establishing a causal relationship between a biological target and a disease state model that closely mimics the human illness state at each stage. Target progression and validation is a strategy that utilizes molecular probes to identify a set of compounds that can modulate the activity of a specific biological target [3]. Target selection can be simplified by employing existing compounds; these 'known' compounds can then be subjected to lead discovery and optimization, resulting in 'new' compounds. Lead discovery involves biologically screening many compounds to identify sets of structurally related compounds with the required biological activity [4]. The lead optimization process starts once a candidate series has been chosen. This phase aims to find a single chemical that can advance into the drug development stage by analyzing structural analogs in a lead series. In lead discovery and optimization processes, many lead series are often found through iterative rounds of experimentation [5], [6]. In many instances, the lead discovery and lead optimization phases overlap since a typical drug discovery program creates multiple groups of related compounds with the potential to find candidates that might move into therapeutic development. Drug development typically follows the lead discovery phase, which concludes when *in vivo* efficacy is established in an appropriate animal model using a molecule with physical and chemical properties consistent with a future clinical trial [6].

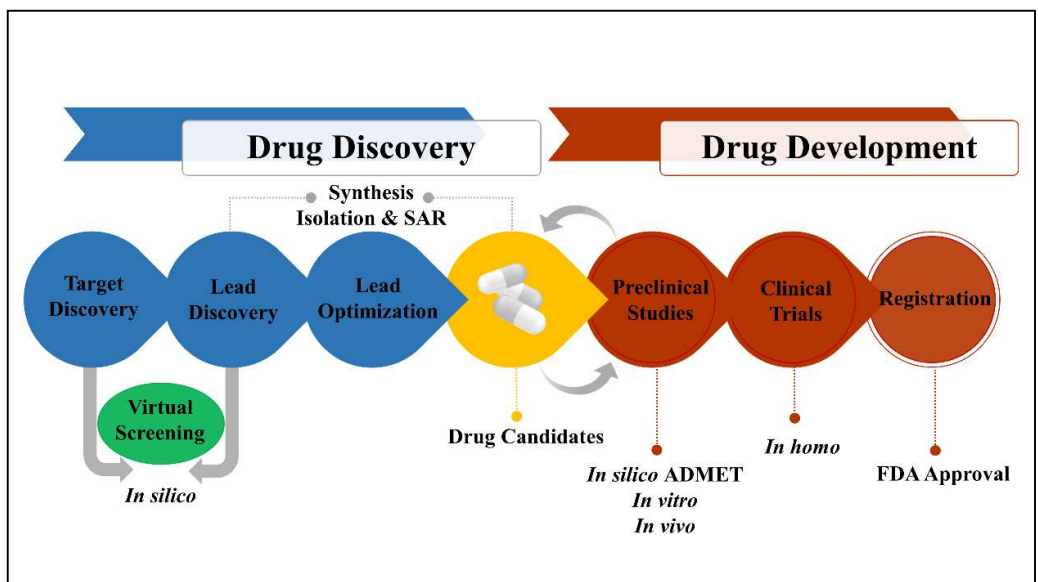

**Figure 1:** An extensive breakdown of the drug discovery and development process

The second main stage, drug development, often begins with identifying a single chemical, which proceeds through numerous studies to support its clearance for sale by the proper regulatory agencies. Preclinical research, clinical trials, and drug registration are three further stages that vary depending on the phase of drug development. As a result of identifying a lead compound (drug candidate), the next step in drug development is preclinical trials [7]. Preclinical studies collect data on a new medicine's safety, effectiveness, and pharmacokinetics in non-human subjects. These studies are conducted both in laboratory settings and through animal models. Researchers employ doses with no limits during animal studies to assess the effects on living organisms, such as mice, and rats, referred to as *in vivo* research. On the other hand, laboratory research performed outside of a living organism, such as in Petri dishes or test tubes, is called *in vitro*, meaning 'in glass'. However, *in silico* assays are biological investigations or testing systems conducted entirely on a computer. The increased interest is foreseen due to developments in computer capacity, behavioral understanding of molecular dynamics, and cellular biology [8]. Phases I, II, III, and IV have the same overarching goals, even if their specific designs vary widely from one candidate to the next. Testing the effectiveness and safety of a novel experimental medication in a small sample of healthy individuals (usually 20-100 persons) is the primary goal of phase I clinical investigations, which aim to determine whether safety margins are acceptable for moving further in the clinical trial process. Findings from the phase determine clinical trial doses in phases II and III of the study. Phase II tests usually involve 100 to 300 patients and are intended to see if the clinical candidate has the necessary biological impact. Due to safety or efficacy problems, most clinical medication candidates fail in phase II research. Phase III enrolls 1,000–5,000 patients, allowing medicine labels and usage guidelines to be developed. The high number of participants, length of time required, and design complexity make Phase III clinical trials the most expensive component of drug research and development. A new drug application is filed to the relevant regulatory body following the completion of phase III trials to get registration and FDA (Food and Drug Administration) approval [6], [9], [10].

The rest of the paper is divided as follows. At first glance, we review the role of artificial intelligence in drug discovery. After that, the detailed steps of drug discovery and design are given in Section 3. Section 4 presents the process of analyzing the compatibility of many molecule structures, known as molecular docking. Then, the process of drug-likeness is overviewed in Section 5, followed by ADMET properties in Section 6. Finally, the binding free energy is calculated in Section 8, and the concluding remarks are discussed.

## 2. Artificial Intelligence in Drug Discovery

The pharmaceutical industry has experienced a significant surge in data digitalization in the past several years. It might be demanding to gather, investigate, and use the knowledge brought about by digitization to tackle complex therapeutic difficulties [11]. This challenge motivates using artificial intelligence, which can handle enormous data sets more efficiently through increased automation. Synthesizing human intelligence, AI uses a wide range of cutting-edge computing resources and computational networks. AI uses systems and software to read and learn from input data to make autonomous decisions to achieve particular goals. Still, it is not a comprehensive threat to replace human physical presence. Machine learning (ML) is a subfield of AI that uses statistical methods to acquire knowledge through or independently of explicit programming [12], [13], [14]. ML employs algorithms to spot patterns in already-organized data [15], [16], [17], [18]. Supervised, unsupervised, and reinforcement learning are the three main sub-types of ML. Classification and regression methods are used to construct predictive models from input and output data while learning is being supervised. Disease diagnosis is an example of supervised ML's subgroup classification output.

In contrast, pharmaceutical efficacy and adverse drug reaction (ADMET: Absorption, Distribution, Metabolism, Excretion, and Toxicity) prediction is an example of supervised ML's subgroup regression output [19]. Feature discovery and data clustering based solely on input data are examples of unsupervised learning techniques [20]. Illness subtype discovery via clustering and disease target discovery with feature-finding approaches possible with unsupervised ML [21]. Reinforcement learning, often known as the 'science of decision-making trial and error', is a method for optimizing performance in a particular setting by implementing decisions. The results of this type of machine learning can be used in various contexts, such as *de novo* drug design for decision-making or experimental design for implementation. Modeling and quantum chemistry help achieve these goals [22] (Figure 2).

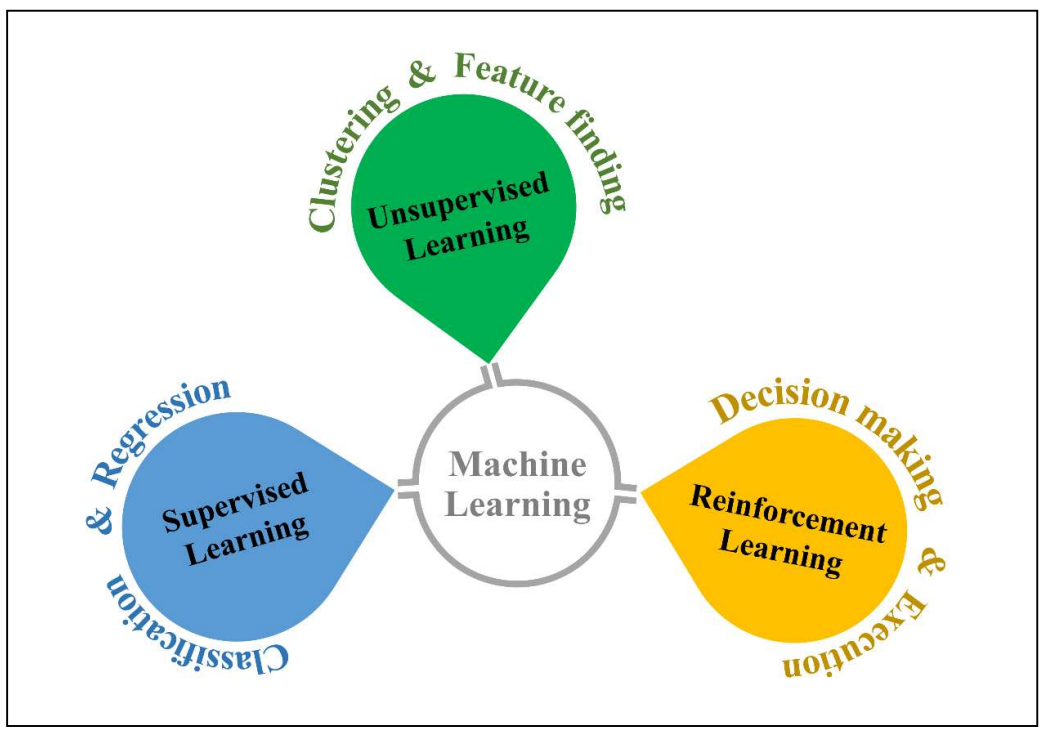

**Figure 2:** Machine learning techniques in drug discovery and development

The artificial neural networks used in deep learning (DL) can learn and adapt from massive experimental data. The advent of big data, along with data mining and algorithm technologies, has the potential to progress the area of personalized medicine based on genetic markers and lead to the discovery or repurposing of current pharmaceuticals that may be more successful, whether taken alone or in combination. The rise of computer power and the expansion of data both contributed to the development of DL [14], [23], [24]. Figure 3 covers the use of AI in the drug development process. This ample chemical space of around $10^{60}$ compounds encourages the creation of a vast array of pharmacological substances. However, due to the lack of cutting-edge technologies, the pharmaceutical development process is lengthy and expensive; AI can overcome this problem. Hit and lead compounds can be found using AI, and the drug target can be validated and the drug structure optimized much more quickly [24], [25], [26].

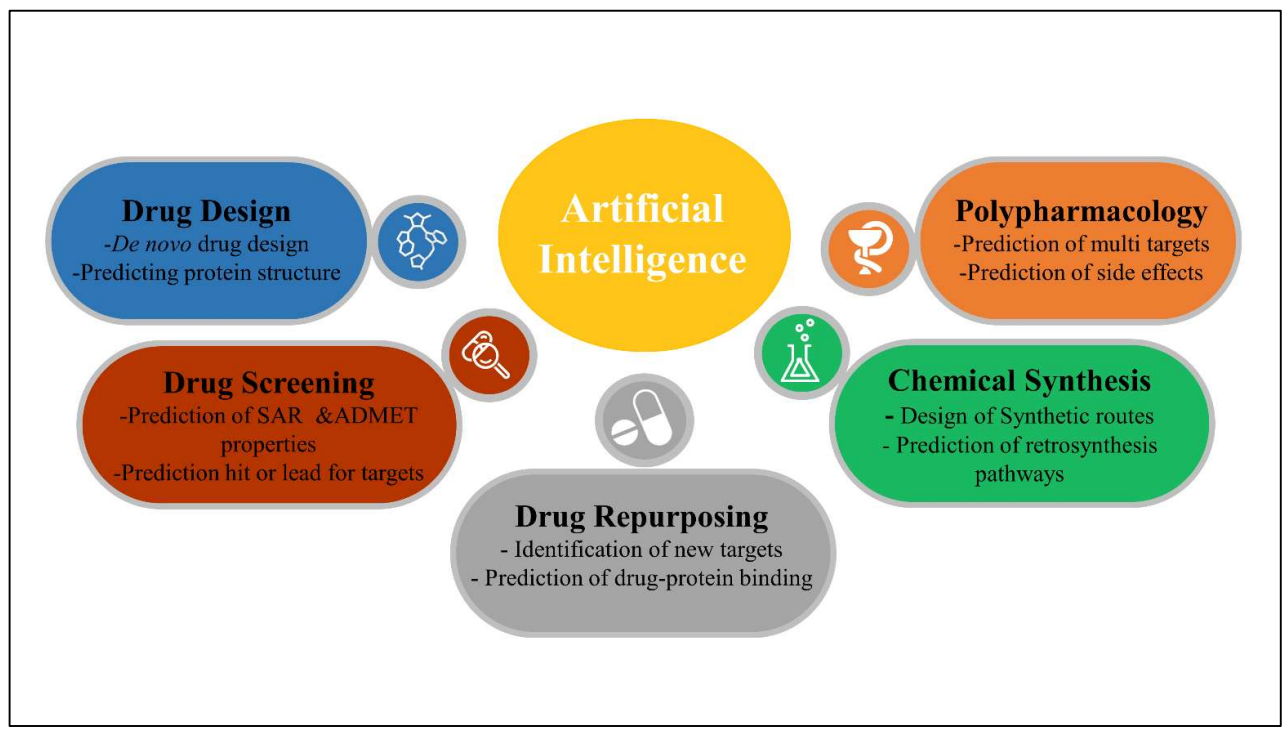

**Figure 3:** Implementation of artificial intelligence in drug discovery

## 3. Drug Discovery and Drug Design

### 3.1 Target discovery

Small molecules, peptides, antibodies, and emerging modalities like short RNAs and cell therapies are the backbone of drug discovery because of their ability to influence the activity of a biological target and, hence, affect the disease state [27]. Two leading causes of drug failure in clinical trials are ineffectiveness and safety concerns. Therefore, target selection and validation are essential in creating a new drug. The term 'target' is broad enough to encompass a wide range of biological entities, including but not limited to proteins, genes, and RNA. One of the essential qualities of a good target is that it is 'druggable' or amenable to therapeutic intervention. Whether the putative drug molecule is small or large, the term 'druggable' refers to a target it can access. The biological reaction can be observed *in vitro* and *in vivo* after binding to the target. The finding of a target with a convincing therapeutic hypothesis that modifying the target would modulate the disease state is still necessary for initiating a drug development program, notwithstanding the current resurgence of phenotypic screens. Target identification and prioritization refer to determining and ranking the essential targets accordingly [28]. Target identification and validation can help us predict the strength of the link between the target and the disease. It allows us to examine whether modulating the target will have unwanted consequences due to the underlying mechanism [29]. A naturally existing (genetic) mutation or a well-designed experimental intervention must be proved to cause the disease to establish that the target is regulated. Increased target identification is primarily attributable to data mining of available biological data. Phenotypic screening, chemo-proteomics, imaging, gene association studies, biomarkers, and transgenic organisms are just a few methods to identify potential therapeutic targets (Figure 4).

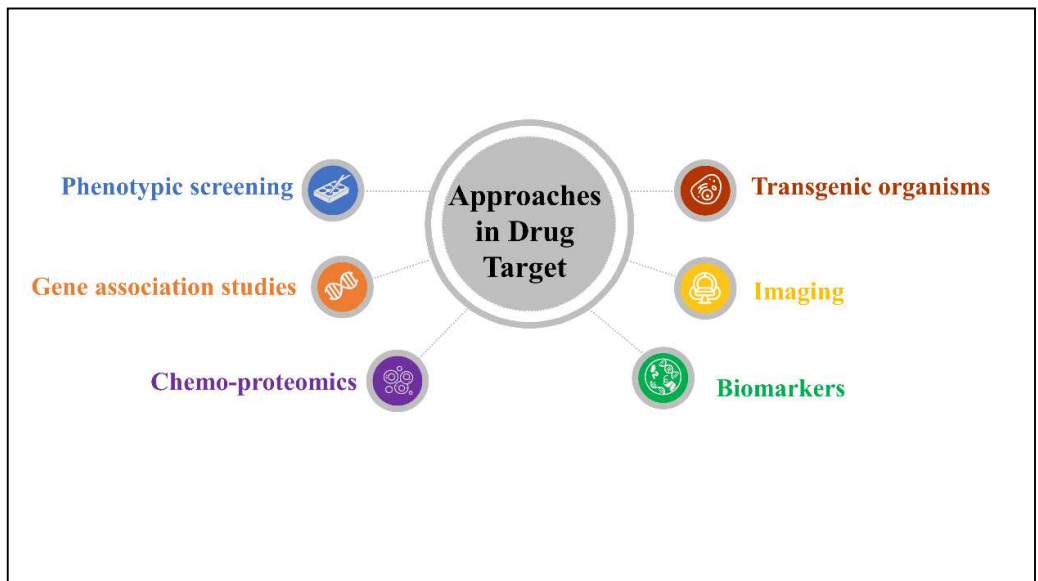

**Figure 4:** Different approaches in drug target identification

Using a computational technique to find, prioritize, and select prospective disease targets is what 'data mining' means in this context [30]. However, to produce statistically effective models that can generate predictions for target identification, these multidimensional data sets need the proper analytical techniques, and this is where ML may be used. Products can be validated using various methods, from *in vitro* instruments and full animal models to modifying a target in patients with the disease. Although clinical trials remain the gold standard for validating targets, early-stage validation using omics technologies and preclinical models enables the efficient allocation of resources toward promising targets [31].

### 3.1.1. Genomics

Genomic data allow researchers to identify mutations or genetic variations associated with diseases, such as single nucleotide polymorphisms (SNPs) [32], copy number variations (CNVs) [33], and other mutations that are implicated in diseases. Techniques such as Genome-Wide Association Studies (GWAS) [34] and whole-genome sequencing have played pivotal roles in identifying genes linked to specific diseases [35], [36]. GWAS studies map genetic loci associated with diseases by scanning thousands of genomes. The studies have proven valuable in uncovering loci associated with complex diseases such as Alzheimer's, cancer, and cardiovascular diseases [34]. CRISPR-Cas9-based screening technologies have also revolutionized functional genomics, enabling researchers to knock out or modify specific genes to determine their role in disease mechanisms [37]. CRISPR screening allows the high-throughput identification of disease-related genes by systematically perturbing the genome. Recent studies have employed CRISPR screens to identify essential cancer genes as potential therapeutic targets [38], [39].

### 3.1.2. Proteomics

Proteomics plays a pivotal role in identifying differentially expressed proteins in diseased versus healthy tissues. It focuses on the large-scale study of proteins, particularly their structures and functions. Proteins are the functional molecules that execute most cellular processes, and changes in protein expression, post-translational modifications, and interactions often indicate disease states [40]. Mass spectrometry (MS) [41] and protein-protein interaction (PPI) networks [42] are critical technologies used in this domain. Mass spectrometry-based proteomics provides high-resolution data on protein abundance, modifications (phosphorylation), and degradation. Integrating proteomic data with genomic and transcriptomic data allows researchers to understand how genetic mutations translate into altered protein functions [43]. PPI networks help elucidate how proteins interact within the cellular environment. Integrating PPI data with disease phenotypes has uncovered networks of dysregulated proteins that could serve as therapeutic targets [44].

### 3.1.3. Transcriptomics

Transcriptomics involves studying the RNA transcripts produced by the genome, providing insights into gene expression patterns under different conditions. RNA sequencing is a widely used tool for transcriptomic analysis and can reveal how diseases affect gene expression. It allows for quantifying RNA transcripts and identifying differentially expressed genes between diseased and healthy states. These genes can then be linked back to the pathways that may be driving the disease, providing targets for therapeutic intervention. Recent studies have integrated transcriptomics with other omics approaches to identify essential regulatory genes that control disease progression, such as transcription factors involved in cancer metastasis [45], [46].

### 3.1.4. Metabolomics

Metabolomics measures the dynamic changes in small molecules (metabolites) within cells, tissues, or organisms. Metabolomic profiling can reveal biochemical changes accompanying disease, providing additional data for target identification. Nuclear magnetic resonance (NMR) [47] and mass spectrometry-based metabolomics are critical for measuring metabolite concentrations. Changes in the metabolome often reflect underlying genetic or protein-level changes and can point to disrupted disease pathways [48], [49], [50].

### 3.1.5. Integrative Multi-Omics Approaches

Combining these omics technologies, referred to as multi-omics integration, allows for a holistic view of disease mechanisms [51]. Researchers can more precisely pinpoint novel drug targets by combining genomic mutations with proteomic changes, transcriptomic dysregulation, and metabolomic shifts [52]. Bioinformatics platforms such as STRING (Search Tool for the Retrieval of Interacting Genes/Proteins) [53], Cytoscape [54], and MetaboAnalyst [55]facilitate the integration of omics data and provide visual representations of complex biological networks. Network-based systems biology approaches have also been increasingly adopted to understand how molecular alterations at different levels (genes, proteins, and metabolites) contribute to disease [56]. Such approaches can predict drug-target interactions and suggest new therapeutic avenues.

### 3.1.6. Protein Structure Prediction
#### 3.1.6.1. Known 3D Protein Structures

Protein structure prediction is critical in understanding protein function and interactions, especially when experimental 3D structures are unavailable. When 3D structures are accessible through resources like the Protein Data Bank (PDB) [57]and the EMDataBank for cryo-electron microscopy structures [58]. These data provide valuable insights into protein function, ligand-binding sites, and molecular interactions, essential for structure-based drug design (SBDD) [59]. The availability of high-resolution structures often determined through X-ray crystallography [60]or NMR spectroscopy [61], enables detailed analyses of protein-ligand interactions, facilitating the rational design of small-molecule inhibitors or modulators. In the context of drug discovery, known protein structures allow for the application of molecular docking and virtual screening techniques, which can be used to predict the binding affinity of various ligands. Furthermore, structural data can inform the design of small molecules with enhanced binding specificity and potency, providing a robust framework for lead optimization and rational drug design [59], [62].

#### 3.1.6.2. Unknown Protein Structures

In cases without an experimentally determined 3D protein structure, computational methods are utilized to predict protein structures based on sequence data. Several approaches have been developed to address this challenge, each offering varying degrees of accuracy depending on the availability of homologous structures and the complexity of the target protein [59].

##### 3.1.6.2.1. Homology Modeling

Homology modeling, also known as comparative modeling, is predicated on the assumption that proteins with similar sequences adopt similar tertiary structures. This method involves identifying a homologous protein with a known structure as a template, followed by alignment of the target sequence to the template to generate a model [63], [64]. Homology modeling is particularly effective when high-sequence similarity exists between the target and the template, typically requiring at least 30% sequence identity for accurate structural prediction. Tools such as SWISS-MODEL [65] and Modeller [66] are widely used.

#### 3.1.6.2.2. Threading (Fold Recognition)

When sequence similarity is insufficient for homology modeling, threading can be employed. Threading methods predict the 3D structure of the target protein by mapping its sequence onto a library of known protein folds, identifying the most likely structural conformation. This approach can be valuable when the target protein lacks close sequence homologs but shares structural features with known proteins [67]. Commonly used tools for threading-based predictions include Phyre2 [68] and I-TASSER [69], which leverage sequence-structure compatibility to produce accurate models for proteins with low sequence identity to known structures.

#### 3.1.6.2.3. AlphaFold Protein Structure Prediction

The recent development of AlphaFold, a deep learning-based algorithm, has significantly advanced the field of protein structure prediction. AlphaFold predicts protein structures directly from amino acid sequences by modeling accurate inter-residue distances and angles. This approach has proven particularly effective for proteins with no homologous structures, providing reliable structural predictions in cases where traditional methods such as homology modeling and threading fall short. AlphaFold is now transforming how unknown protein structures are modeled, making it an indispensable tool for studying protein function and facilitating drug discovery [70], [71].

### 3.2. Lead discovery

After target identification, the lead discovery step of the small molecules' translational route involves the search for compounds that can interact with the target. Pharmaceutical chemists look for substances that might interact with the target. During lead discovery, scientists look for a drug-like biological or small chemical treatment that can be put through preclinical studies, clinical trials, and, if everything goes well, into the market. The term 'development candidate' describes this class of drugs [72].

### 3.3. Hit identification

Finding 'hits' or molecules that interact with the target despite being poorly optimized, is the first step in the lead discovery process. To determine which functions are most effectively modified, this method considers tiny molecules as hits for target binding [72]. One of the essential milestones in preclinical drug discovery is the identification of drug-target interactions. A drug's therapeutic effects depend on its interaction with its intended target, while off-target protein interactions can lead to unwanted side effects and drug repositioning [73]. Furthermore, due to the limited number of currently available drugs, it is not easy to experimentally examine the entire chemical space of chemicals for druggable target proteins. Since the conceivable organic molecules may be enumerated computationally in chemical space, it is possible to identify novel and high-quality molecules [74].

Nevertheless, information about chemicals, medications, proteins, and their biological activity rapidly accumulates, making it possible for data-driven computation models to find hits in a vast chemical space. Several computational models have been developed to discover drug-target interactions and estimate binding affinities, which has the added benefit of delivering innovative drug candidates during the early stages of drug development [75]. Moreover, AI systems can reduce attrition rates and research and development (R&D) spending by decreasing the number of produced compounds tested in *in vitro* or *in vivo* systems. Validated AI techniques can be utilized to improve drug development success rates, while the currently under-development AI approaches must be verified before being applied to the medication development process. The most crucial phase of drug discovery is the stage in which the desired molecules are synthesized. To that end, AI is helpful since it can rank molecules according to how simple they are to synthesize and provide practical tools for selecting the most effective synthetic method [76], [77]. Several different types of screening paradigms can be used to locate hit chemicals.

### 3.3.1. Drug repurposing

Drug repurposing aims to discover new uses for already approved pharmaceuticals by investigating their potential for new effects or targets. It is considered that drug repurposing offers significant advantages over the typical method of drug development (*de novo* drug discovery) that involves hunting for a novel active ingredient with time and money-saving effectiveness. Drug repositioning potentials could be developed more rapidly, reducing development risks due to applying existing drug knowledge [78], [79], [80], [81], [82]. Other terms for this practice include therapeutic switching, drug reuse, drug recycling, drug re-tasking, and drug reprofiling. Typical drug repositioning research focuses on finding drugs with similar effects [83] and mechanisms of action, looking at shared properties such as chemical structures and adverse effects [73], establishing links between diseases and medications [84], or selecting new targets from the present pharmacopeia to screen for novel medication indications [85]. Figure 5 shows how drug repurposing assists in the discovery of novel drugs.

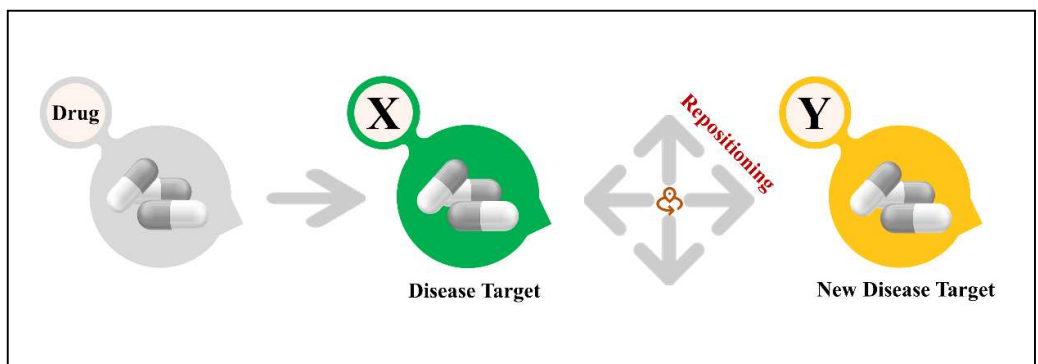

**Figure 5:** Drug repurposing: Drug approved for X(Green), finding new target disease named Y (Yellow) from an old existing drug

AI makes the process of repurposing drugs more appealing and practical. Using an already approved treatment for a different disease is favorable since it allows the drug to move immediately from Phase I clinical trials and toxicology testing to Phase II studies for an additional indication [24], [86], [87]. Although repurposing drugs is nothing new, the term 'drug repurposing' as a technique in R&D for new therapies did not appear in the literature until 2004 [78]. This strategy, however, may have been put into action in the late 1990s when thalidomide was reclassified. One of the worst medical tragedies ever occurred in the 1960s, when this drug was widely administered as an antiemetic for pregnant women. In the same decade that it was discovered that a racemic mixture of thalidomide caused birth abnormalities, the drug was taken off the market [88].

Nevertheless, cytosine, found in high amounts in leprosy patients, is selectively inhibited by thalidomide, which is crucial in this process. Based on these investigations, the FDA approved the medication, in July 1998, for treating erythema nodosum leprosum, a severe and crippling illness related to leprosy [89] [88]. In 2000, thalidomide was once more repurposed for treating refractory multiple myeloma; its usage is controlled in Brazil [79], [90]. *In silico* drug repositioning has become more practical due to the rapid expansion of cancer genome data and the pharmaceutical industry's gradual sharing of internal high-throughput screening (HTS) data. More than 200 non-cancer medications have been demonstrated to possess anticancer properties. Propranolol, frequently prescribed to control blood pressure, has recently been shown to be beneficial in reducing tumors when used with other cancer treatment methods [91].

### 3.3.2. High-throughput screening (HTS)

Drug discovery methodologies have significantly modified and included numerous new ideas during the past few decades. HTS emerged from efficiently screening manageable library sizes. Whether using a cell-based assay or a more complicated assay method that relies on many targets and, as a result, requires additional tests to confirm the site of action of drugs, screening is the process of assessing a large number of compounds for their activity against a therapeutic target [92]. This screening strategy uses advanced laboratory automation but does not assume (based on prior information) the type of chemotype that will be most effective at the target protein. High throughput is developed and implemented to find compounds interacting with the therapeutic target.

Chemical programs are carried out to boost the potency. The molecule's selectivity and physiochemical characteristics support the premise that a therapeutic target intervention will benefit the illness state. Research in the pharmaceutical business and, increasingly, in academic institutions is focusing on this sequence of activities to identify potential molecules for clinical development. To find targets, collect compound libraries, and set up the necessary infrastructure to screen those compounds, pharmaceutical companies have developed massive organizations to identify hit molecules from HTS and optimize those screening 'hits' into clinical prospects [29], [93]. Due to the high cost of HTS screens, the HTS assay needs to be optimized and reduced in size. The standard number of wells on a screening plate is 1536 [94], indicating that HTS enables a rapid evaluation of several chemicals, typically highly automated. The 'hit compounds' activity will be examined at various concentrations, or a dosage response, after they have been identified. The design and synthesis of extensive and varied libraries of small organic compounds via combinatorial chemistry have cost the pharmaceutical industry much. An ever-growing amount of data is produced when these chemical libraries are evaluated using HTS techniques against potential therapeutic targets. This data must be handled smartly and highly automated to find and select novel hits and lead candidates [95].

Currently, chemoinformatic methods help medicinal chemists and molecular pharmacologists examine and organize data and comprehend connections between observed biological data and drug attributes. The effectiveness of HTS depends on the choice of the most probable lead compounds, the prompt and systematic removal of suspected 'false positives', and the detection of 'false negatives'. The properties of the compounds selected for different lead-finding processes should ideally be druglike in several pharmacokinetic, physicochemical, and structural criteria. Chemoinformatic approaches play a significant role in identifying and implementing these factors to reduce the size of data sets because *in vitro* and *in vivo* screening cannot handle the generally high number of primary HTS hits [96], [97], [98]. Several diseases, such as multiple myeloma, glioblastoma, and pediatric tumors, have benefited from recent HTS developments in the design of customized therapy. Several drugs have been studied for their potential to provide targeted treatment, including the mTOR inhibitor hemisodium, the ALK inhibitor ceritinib, and the PLK1 inhibitor BI2536; the proteasome inhibitor bortezomib; the Bcl-2 inhibitor ABT-263; and the mTOR inhibitor AZD-8055 [99], [100], [101].

### 3.3.3. Virtual screening (VS)

To reduce the time, money, and resources needed to find novel compounds, virtual screening techniques have revolutionized the discovery of new compounds with specific bioactivity by analyzing massive structural libraries against a bioreceptor or biological system [102]. VS can be implemented by docking a known medication into many diverse target structures or by docking a database of licensed drugs into a single intended specialized target. These techniques use iterative and hierarchical steps to narrow the search space to only those compounds that meet specific criteria for their pharmacokinetics, pharmacodynamics, and physicochemical properties. Hit compounds pass through every filter of the VS, and to verify their biological activity, they must undergo experimental testing in the lab. A virtual screening process includes five fundamental tasks; the first four steps are performed computationally, while the last is experimental (Figure 6). The first is library preparation, which entails accumulating chemical structures of substances (ligands and receptors), converting files to usable formats like PDBQT (Protein Data Bank, Partial Charge (Q), & Atom Type (T)), SMILES (simplified molecular-input line-entry system), SDF (structure data file), and MOL2 (MDL Mol file) [103], [104], [105], [106]; conformers creation, stereochemical correction, solvent and receptor-associated binding substances removal, polar hydrogens and charges addition are among the processes [107]. The data is pre-processed and filtered in the second step, and drug-likeness properties are examined in the third step. The fourth stage employs computational methods [108], docking studies, molecular dynamics (MD) simulation, and molecular mechanics calculation to filter the desired chemicals. However, the last phase involves experimental validation employing *in vitro* and *in vivo* assays, such as cell lines and enzymatic inhibition studies [109], [110]. In the early stages of drug development, VS approaches are often used to expand the initial library with active compounds, much as HTS. Compared to HTS, VS can handle thousands of compounds quickly, reducing the number of compounds that must be manufactured, purchased, and analyzed.

Additionally, VS can be used to create virtual libraries of substances, increasing the chemical space. While a VS strategy occasionally produces highly active compounds, its primary objective may be to generate structurally diverse lead compounds that can be optimized further during the hit-to-lead and lead-optimization phases [111]. Virtual screening strategies have been applied using a variety of computer methodologies that have been developed over time and draw on an understanding of molecular modeling [112], [113], [114], [115], [116], probability and statistics [117], [118], [119], [120], [121], and artificial intelligence [122], [123], [124], [125], [126]. The success of discovering novel bioactive chemicals increases when these techniques are combined with experimental procedures [127], [128], [129], [130].

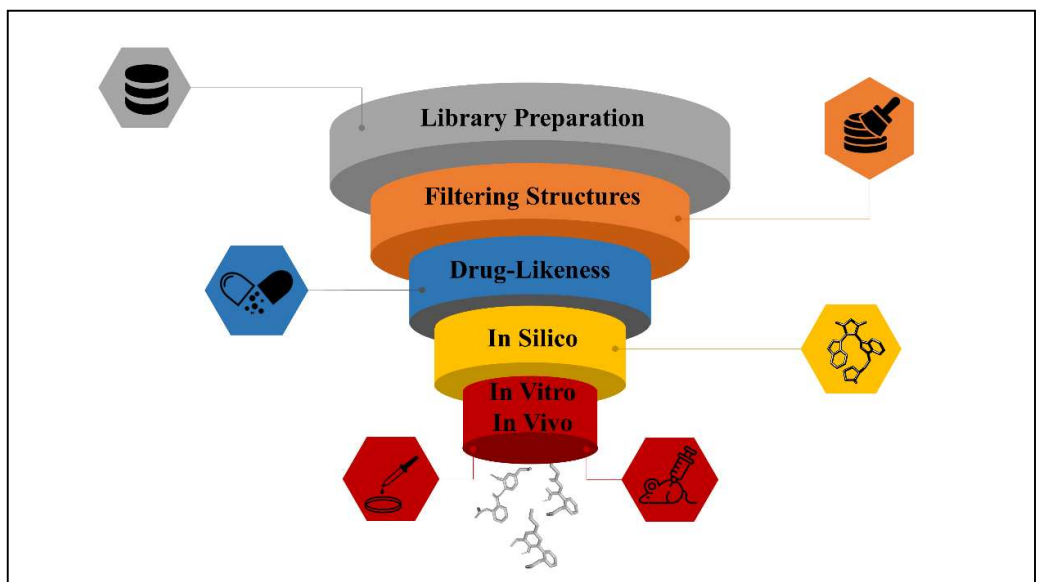

**Figure 6:** Steps involved in the virtual screening process

The methods used for virtual screening can be broken down into two distinct groups. The first is the Ligand-Based Virtual Screening (LBVS) method, which finds promising compounds based on how closely they resemble already-approved drugs. Secondly, the Receptor-Based Virtual Screening (RBVS) technique emphasizes the complementary nature of the desired compounds with the target protein's binding site [108].

### 3.3.3.1. Ligand-based virtual screening

LBVS, researchers compare compounds in a library to reference compounds that are effective against a target or possess the desired characteristics. Johnson and Maggiora initially proposed similar properties [131], according to which similar compounds have similar properties and serve as the foundation for various strategies. As a result, substances with a high degree of resemblance to reference substances are more likely to exhibit similar behavior or function via equivalent mechanisms and produce comparable effects. Since medications can have many pharmacological effects, similarity has been widely used in lead identification and optimization [132]. Different methodologies utilize various similarity measurements when comparing two compounds because similarity is subjective. Compared to structure-based methods, the computational cost of ligand-based VS approaches is lower because no macromolecules are required. This aspect means they are most useful at the outset of the VS procedure when the number of chemicals in the initial library is the biggest. The LBVS technique only utilizes assessments of the basic properties of the compound structure, for example, physicochemical, topological, structural, and electronic aspects of a molecule that are connected to its molecular activity, a group of chemicals with empirically demonstrated biological action as a starting point [133], [134], [135]. The LBVS approach uses various computational methods [108], [136]. The three most common techniques are Ligand-based Pharmacophores, which are constructed by superimposing a set of active compounds to determine the standard chemical features that promote their activity [137]. Similarity searching involves collecting molecules from a database with user-defined query substructures comparable to active substances [138]. Machine learning methods are commonly employed in quantifying structure-activity relationships (QSAR) [139].

### 3.3.3.2. Structure-based virtual screening

The structure-based virtual screening (SBVS) method designs and finds new bioactive compounds by using knowledge about the ligand's molecular recognition in the bioreceptor structure as a starting point. This information also includes the bioreceptor's geometry, the molecule's surface charge, intermolecular interactions, and the chemical composition of the residue at the binding site [130], [140], [141]. These techniques require the receptor's 3D structure to be fully understood and, ideally, combined with the bioactive substance. The 3D structure determines the bioactive ligands' structural conformation and molecular binding location. Molecular docking, molecular dynamics simulation, and structure-based pharmacophore modeling are examples of computational techniques used in the SBVS methodology [108], [142]. Public databases like the RCSB Protein Data currently contain more than two million 3D structures of molecular targets that have been determined. SBVS techniques investigate information on the target protein structure when choosing compounds expected to interact favorably. Due to its knowledge-based nature, SBVS places a high value on the quantity and quality of information accessible on the system under study. Regardless of the protein type selected as a molecular target, it is essential to consider the target's draggability, the selection of the most pertinent shape, the receptor's flexibility, the correct allocation of protonation states, and the existence of molecules of water in the binding site [143], [144]. The various steps involved in the LBVS and SBVS are briefly shown in Figure 7.

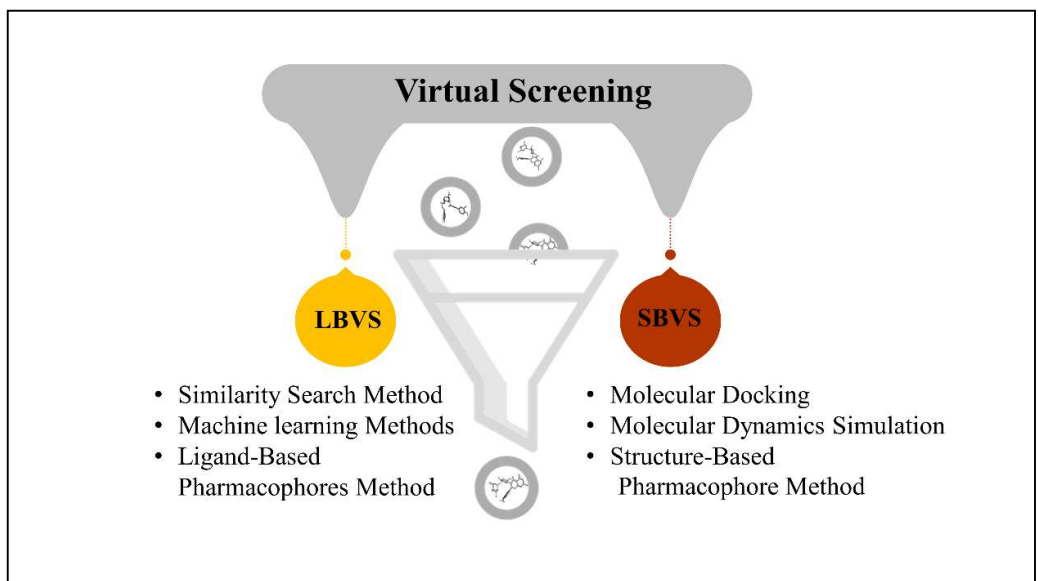

**Figure 7:** Types of virtual screening: ligand-based virtual screening (LBVS) and structure-based virtual screening (SBVS), as well as corresponding computational methods

### 3.3.4. Network Pharmacology

Network pharmacology is a cutting-edge approach to drug discovery that moves away from the traditional idea of 'one drug, one target'. Instead, it focuses on understanding the complexity of human diseases at a broader systems level 'multi-target, multi-drug' model. By combining systems biology, computational modeling, and experimental methods, network pharmacology aims to map out the complex interactions in biological networks, such as protein interactions, gene regulation, and metabolic pathways. This approach helps scientists find and affect multiple targets within a network, which is especially useful for diseases like cancer, neurodegenerative disorders, and metabolic syndromes, where disruptions across many pathways drive the disease forward [145], [146].

A key concept in network pharmacology is identifying 'network hubs' or important nodes in these biological networks that control several pathways. Targeting these hubs allows drugs to have a more powerful effect compared to traditional single-target drugs, which may only address one part of a disease. For instance, in cancer treatment, the interconnected nature of signaling pathways can cause other pathways to compensate when a single protein is blocked. Network pharmacology addresses this issue by developing multi-target therapies that block several critical pathways simultaneously, lowering the chance of resistance and improving treatment results [147].

Recent advances in computational tools have greatly improved the use of network pharmacology. Tools like STRING, Cytoscape, and STITCH help build and visualize complex interaction networks, making identifying key targets and druggable points easier. These tools combine data from multiple omics layers, such as genomics, proteomics, and metabolomics, to provide a complete picture of the molecular mechanisms behind diseases (as described in Section 3.1.5. Integrative Multi-Omics Approaches) [54]. New algorithms for network-based drug repositioning allow researchers to find new uses for existing drugs by analyzing their effects across different biological networks. These discoveries speed up the drug discovery process and reduce the cost and time needed to develop new therapies [148], [149].

The benefits of network pharmacology are significant. It gives a more complete understanding of how diseases work, enabling the creation of drugs that can target multiple areas with greater precision and effectiveness. By considering the interconnected nature of biological systems, network pharmacology reduces the risk of side effects and increases the likelihood of finding drugs with safer profiles. It also offers a strong framework for drug repurposing, helping researchers find new treatments for existing drugs by examining how they influence different molecular networks. As the field evolves, incorporating AI and machine learning is expected to further speed up the discovery of multi-target drug candidates, leading to more personalized and effective treatments in the future [150], [151], [152].

## 3.4. Hit to lead

Lead generation refers to the first phase of drug development. Once a prospective pharmacological target has been discovered, the next step in the research process is identifying molecules that can engage with the target to produce the desired biological effects. A hit compound is a molecule that reaches the required activity level in a screening assay. The hit discovery and hit-to-lead selection processes rely heavily on developing pharmacologically relevant screening assays. By refining the screening parameters, the most potential molecules in a pool of hits are singled out as potential lead compounds [72], [153], [154]. In a typical hit-to-lead process, the verified hit is chemically modified to increase its affinity for the target and transform it into a lead compound. These optimization phases can be finished using a trial-and-error approach, and it typically takes several cycles to find the suitable affinity. In addition to off-target effects, physicochemical properties and ADME features can be screened using the secondary assays used in the lead selection process. This discovery process step aims to improve each successful series to create more powerful and selective drugs with pharmacokinetic (PK) characteristics enough to determine their effectiveness in any accessible *in vivo* models. The critical factors for hits obtained using various hit-finding algorithms may differ remarkably [155]. It is standard procedure to conduct in-depth structure-activity relationships (SAR) studies around each central chemical structure and then to use these data to ascertain the activity and selectivity of the molecules in question. This process must be carried out systematically when structural knowledge of the target is available.

Structure-based drug design methods, including molecular modeling and techniques like X-ray crystallography and nuclear magnetic resonance, can be utilized to build the SAR more rapidly and precisely. This action typically leads to discovering new binding sites on specific proteins [29], [72], [153], [156]. There are a variety of computer-based tactics designed to circumvent hit-to-lead ratios. For instance, by utilizing accessible experimental data and derived descriptors, ligand-based approaches like the Quantitative Structure-Activity Relationship (QSAR) can be performed to improve a sequence of compounds. To optimize a set of compounds, for instance, ligand-based techniques like the QSAR can use the existing experimental observations and derived descriptors [157]. By using the binding interactions of the confirmed fragment as an initial point, *de novo* design methods can also be used to sample the cavity and generate potential optimum molecules [158], [159].

### 3.4.1. Quantitative structure-activity relationship

The hit-to-lead optimization technique used QSAR analysis to identify possible lead compounds from hit analogs with the prediction of bioactivity analogs [160]. It is a computer program used to calculate the strength of a correlation between the chemical structures of a set of substances and an observed chemical or biological reaction. Comparable structural or physiochemical properties lead to relative activity; this is the core idea behind the QSAR [161], [162]. QSAR models provide a mathematical explanation of how a ligand's structural characteristics influence the behavior of a target after it binds to it. Electronic, hydrophobic, steric, and sub-structural molecular characteristics may be applied to build QSAR models [163]. The first step is to identify a set of chemical building blocks or lead molecules that display the target biological activity. There is a robust quantitative relationship between the physicochemical properties of the active compounds and their biological activity. The developed QSAR model is then used to enhance the biological activity of the active compounds. Next, the anticipated chemicals are tested in the lab to see if they have the expected action. As a result, the QSAR method can be used as a map to locate alterations to a chemical that result in a noticeable increase in activity. QSAR approach has advanced primarily regarding the variety of molecular descriptors utilized and their relationship to activity [164].

To construct a reliable QSAR model, a set of sequential processes is required [72], [164], [165], [166] (Figure 8): (i) Find ligands that exhibit the desired biological action (a minimum of 20 active chemicals) at experimentally observed levels. Though they should have a good chemical variety to display a wide range of activities; (ii) Choose and identify the molecular descriptors that correspond to the different construction and physicochemical characteristics of the molecules under examination; (iii) Analyse the data set to find chemical descriptors that correlate with biological activity and provide an explanation for the observed variance in activity; (iv) Examine the QSAR model's statistical applicability and predictivity performance.

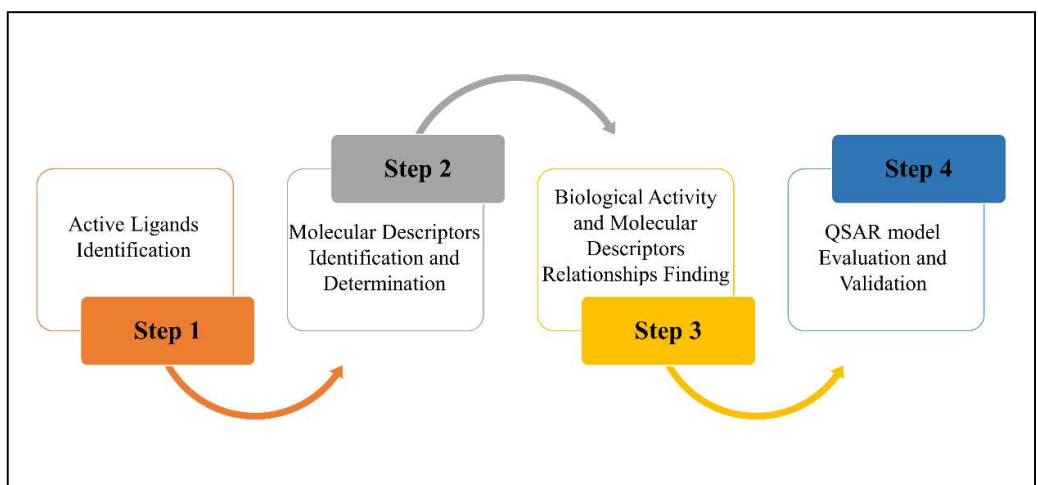

**Figure 8**: Standard QSAR technique workflow

### 3.4.2. *De novo* drug design

*De novo* design tools use predetermined collections of substructures and rules guiding their linking to automatically create compounds 'from scratch' within a binding site with a known 3D structure of the receptor. Target receptor information is provided, but no leads can interact with it. Molecular modeling methods determine the lead target complexes' structural properties and modify the lead [158], [159]. In contrast to traditional virtual screening, which only tests chemicals readily available on the market, *de novo* approaches theoretically allow the study of much greater chemical space. Despite having a similar principle, *de novo* tools can be identified by their 'algorithms, convergence criteria, ranking/scoring functions, and branch-pruning' techniques. This methodological approach, created in the early 1990s, was initially promising but no longer extensively used by chemo-informaticians [153], [167], [168], [169], [170], [171], [172], [173]. A *de novo* lead molecule is made by inserting building blocks or fragments into the active area of the target protein and then connecting them using chemical rules, bridges, or linkers. However, the *de novo* evolution method starts with a single scaffold or building block and inserts it into the active site's crucial interaction region, followed by the addition of additional fragments or functional groups (such as alcohols, amines, single rings, esters, and simple hydrocarbons) to increase binding affinity to the specific protein (receptor) [174], [175], [176], [177], [178] (Figure 9). Indeed, *de novo* approaches frequently have several significant disadvantages that limit their use to hypothetical cases. The key disadvantages include the fact that the developed compounds may have poor physicochemical properties, low accuracy in predicting affinity, and concerns about the synthetic flexibility of the compounds [159]. Deep learning models drew on their rich knowledge and generative abilities to generate a new structure with beneficial properties. Deep learning models act as autoencoders in the *de novo* drug discovery, providing a valid structure for new chemicals (NCE). Therefore, an auto-encoder integrating with a multilayer perceptron profiler is valuable for creating NCEs with specified physicochemical properties [72].

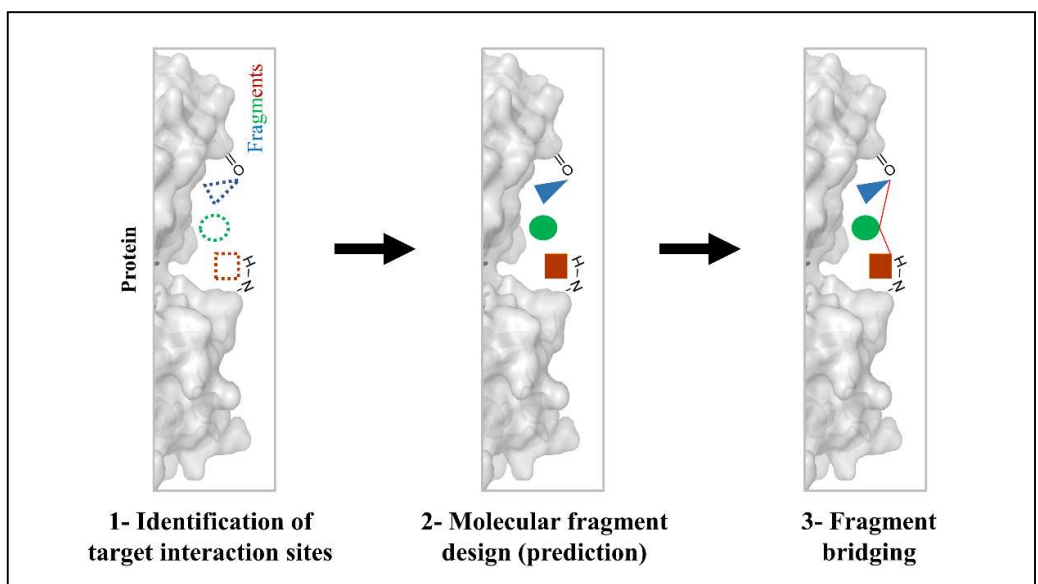

**Figure 9:** Principles of *de novo* drug design

### 3.4.3. Fragment-Based Drug Design (FBDD)

FBDD is an alternative approach in drug research that focuses on finding small molecular fragments, typically under 300 Daltons, which weakly bind to target proteins. FBDD aims to refine these fragments into larger, high-affinity compounds by merging, expanding, or connecting them to form drug-like molecules. Compared to traditional high-throughput screening, FBDD provides several advantages, such as more efficient exploration of chemical space, higher success in targeting challenging proteins, and the ability to design leads using structural data. Identifying fragments involves using highly sensitive techniques to detect weak binding interactions [179]. X-ray Crystallography is a common method, revealing the three-dimensional structure of the fragment bound to the protein, allowing visualization of atomic-level interactions. NMR Spectroscopy can detect small fragments, even with weak binding, by analyzing changes in the magnetic properties of atomic nuclei. Surface Plasmon Resonance (SPR) measures real-time binding affinities of small fragments to proteins by detecting changes in refractive index on a biosensor surface [180].

In FBDD, computational techniques are essential for identifying and refining fragments. Virtual screening is frequently employed to assess fragment libraries by simulating the docking of fragments into the binding sites of target proteins, ranking their positions based on predicted binding energies. This significantly reduces the reliance on extensive experimental testing (refer to Section 3.3.3 Virtual Screening). Molecular docking further helps to forecast the binding modes of fragments, modeling their interactions with the target to identify the best orientations and binding affinities (see Section 4. Molecular Docking). Moreover, molecular dynamics (MD) simulations offer insights into the stability of fragment-protein interactions over time by analyzing flexibility and revealing key interactions that static docking models may miss (Section 7. Molecular Dynamics Simulations). To refine these predictions, free energy calculations evaluate binding affinities, helping to prioritize fragments for further development (refer to Section 8. Binding Free Energy Calculation). Together, these computational tools streamline selecting and optimizing fragments, accelerating the drug discovery process in FBDD [181], [182].

Once a fragment is identified through experimental screening or computational docking, it is optimized to improve its binding affinity and drug-like characteristics. Various strategies are employed for this optimization. One approach, fragment growing, involves adding chemical groups to the fragment in a systematic way, which increases its size and the interaction surface within the binding site. Structural data, such as from X-ray crystallography or docking studies, often guides this process. Another method, fragment linking, happens when two fragments bind to adjacent areas of the target protein; these fragments are then chemically connected to form a larger molecule with better binding properties, ensuring proper alignment and interactions with key residues. Fragment merging is another technique, where fragments that bind to overlapping regions of the protein's active site are combined into a more potent compound. This approach typically uses structure-based drug design SBDD techniques to retain favorable binding interactions and enhance the drug-like properties of the molecule [181], [183].

### 3.5. Lead optimization

The most promising fragment hits are identified for additional medicinal chemistry study in the lead optimization step, facilitating the essential steps in the drug development process. By reducing the amount of structural change made and doing away with the adverse effects of the currently active analogs, lead optimization aims to produce a better and safer scaffold [72]. The lead optimization phase is complete when the primary goals have been met, and the molecule is ready for final characterization before being labeled as a preclinical contender. Even at this point, discovery work is ongoing. The researcher has to keep searching for comparable series and doing synthetic research in case the product undergoing further preclinical or clinical characterization does not pan out. Depending on the company, the lead optimization phase may incorporate a subset of the various activities that comprise enhanced characterization. Once enough information has been collected about the molecule; a target candidate profile can be developed; this profile, combined with data on the molecule's toxicity and manufacturing and control concerns, will form the basis for regulatory submission to begin human administration. The initial screening of 200,000 to $>10^6$ compounds and the subsequent hit-to-lead and lead optimization programs may occur for each project in the industry. Numerous compounds, typically from different chemical series, are tested to narrow the field to one or two candidate molecules. Due to the expensive cost of an extensive HTS or the usage of chemicals produced from a structure-based method, targeted screens are more common in academia [184]. Although many small molecule projects in the industry have the potential to become candidates, only about 10% succeed. The candidate molecules may fail at multiple phases, including (i) being unable to set up a dependable assay; (ii) the HTS yielded no hits that could be developed; (iii) secondary or native tissue tests of chemicals do not produce the intended results; (iv) substances are lethal at the preclinical stage (*in vitro* and *in vivo*); (v) compounds have unwanted side effects that are difficult to filter out or distinguish from the target's method of action; (vi) being unable to attain a decent pharmacokinetic or pharmacodynamic profile following the dosage regimen needed in human, for instance, if this calls for a once-daily tablet, the chemical must have an appropriate *in vivo* half-life to accomplish the task; and (vii) substances with a central nervous system target cannot penetrate the blood-brain barrier [29], [185].

# 4. Molecular Docking

The process of analyzing the compatibility of many molecule structures is known as molecular docking. To better understand how small compounds behave at the binding site of target proteins and to shed light on fundamental biological processes, the molecular docking technique may mimic the interaction between a compound and a receptor at the atomic level [186].

## 4.1. The fundamental concepts of molecular docking

The docking method involves predicting how a ligand will fold within a binding site and then positioning the ligand in that folded state. There are two main aims of docking research: structural modeling and activity prediction [187]. Figure 10a depicts the 'lock-and-key model', which uses rigorous docking of receptors and ligands to identify the optimal orientation of the 'key' to unlock the 'lock'. This theory emphasizes the relevance of geometric complementarity [188]. Fisher initially proposed the lock-and-key model idea in 1894 to characterize the conceptual receptor-ligand interaction theory, which stated that ligands and receptors could identify each other through geometric compatibility and energy matching [189]. The drug and receptor molecules are viewed as immovable locks and keys in the lock-and-key model. Before and following binding, the three-dimensional structure and orientation of the drug and receptor undergo modest modifications, which this model can explain. However, more incredible conformational changes cannot be explained by this model. However, to account for the constraints of the lock-and-key model as well as the changes in enzyme conformation imposed by the substrate during the interaction between the enzyme and the ligand (substrate), Koshland created the induced fit model in 1958 [190], [191] (Figure 10b).

In the induced fit model, both the ligand and the protein are malleable structures during molecular docking, as shown by the fact that the active site spatial configuration of the protein changes due to contact with the ligand [192]. Geometric complementarity underlies the principles of energy complementarity and pre-organization, guaranteeing that receptors and ligands will reach the most stable structure while reducing free energy [193]. Applying the induced fit model to the interaction between a drug molecule and a receptor has improved the docking findings by treating ligands and receptors as rigid structures [194].

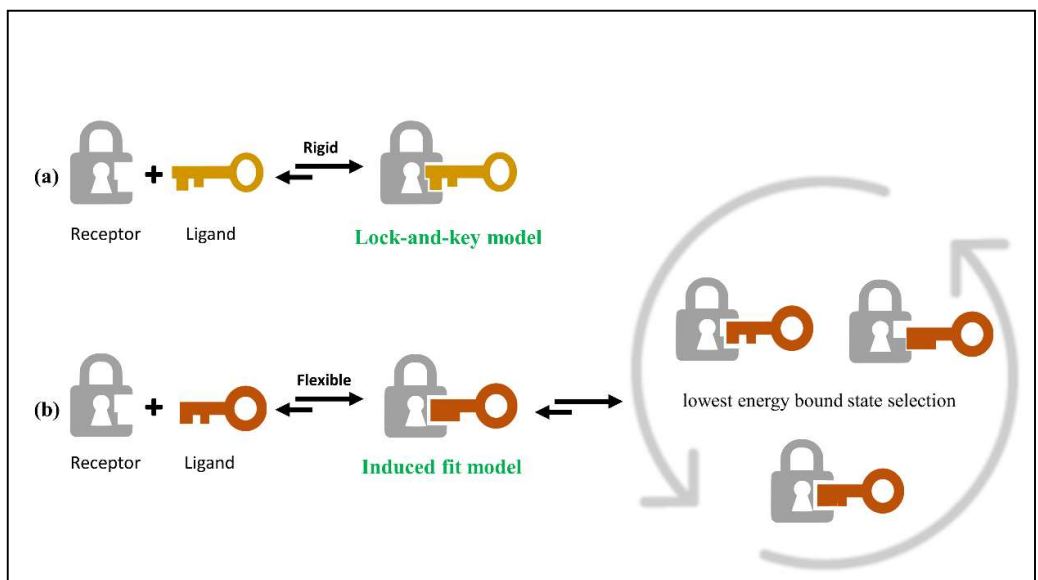

**Figure 10:** Molecular docking models

### 4.2. Molecular docking types

#### 4.2.1. Flexible docking

Flexible docking allows the ligand and receptor conformations to be flexible during the docking calculation. This docking simulation is widely used to rigorously analyze the identification of two molecules because it is so realistic and close to the actual docking situation. However, due to the geometric growth of factors with the number of atoms in the system, the flexible docking approach is computationally time - consuming and costly, placing significant demands on both software and hardware systems [192]. Popular molecular docking programs include FlexX [195]. Many methods have been developed, some depending on conformational choices and others on the induced-fit binding concept. The potential energy surface is very coordinate-dependent due to the considerable increase in the degrees of freedom introduced by flexible docking. Therefore, there is an increase in the computing effort needed to carry out a docking calculation. It is essential to balance speed and accuracy. Thus, sampling and scoring should be adjusted accordingly. The speed of docking computations influences a virtual screening campaign involving millions of chemicals. As a result, the new algorithm's development has continued to advance, allowing it to thoroughly search the phase space without reducing velocity [196].

#### 4.2.2. Semi-flexible docking

The receptor's conformation is held constant during the semi-flexible docking calculation process. In contrast, the ligand's conformation can vary within a range determined by particular parameters, such as the bond length and angle of some non-critical components. Docking simulations involving tiny molecules and biomacromolecules, including proteins, nucleic acids, and enzymes, have become increasingly popular due to this method's ability to compute and predict models [187]. Reportedly, FlexX, Dock, and AutoDock are the most popular semi-flexible docking programs [192], [197], [198].

### 4.2.3. Rigid docking

Rigid Docking determines the strength of a binding interaction between two molecules. In this procedure, the molecules' conformations are not changed; instead, they switch positions in three-dimensional space [199]. This approximate approach is comparable to the 'lock-key' binding model [196]. This docking simulation assumes that the ligand and receptor will always be in the same spatial configuration. This docking method is the most practical because it requires the fewest and least extensive calculations. Therefore, it may be used to investigate relatively large structures, such as protein-protein and protein-nucleic acid complexes, which are essential parts of docking systems [192].

### 4.3. Search methods and ligand flexibility

Three fundamental categories exist for the treatment of ligand flexibility, and some algorithms combine many of these techniques [200]. In a systematic search, all potential values for each coordinate are explored in a combinatorial method by first assigning those coordinates to grid values. As the number of degrees of freedom rises, so does the needed number of evaluations. The algorithm is prevented from randomly sampling from regions known to yield inaccurate results thanks to the introduction of termination conditions. The 'anchor-and-grow' algorithm or incremental construction algorithm are examples of systematic searches [200], [201]. Random search (stochastic search): stochastic search algorithms randomly alter the system, typically modifying one degree of freedom at a time. The convergence uncertainty of stochastic searches is one of the main issues. Multiple, separate runs can be made to enhance convergence. Evolutionary algorithms and Monte Carlo (MC) techniques are examples of stochastic searches [200]. Deterministic search (simulation methods): in deterministic searches, the initial state determines the action that may be done to produce the next state, which typically must have the same energy as the first state or less. Deterministic searches with the same initial system (including all degrees of freedom) and parameters will always yield the same solution state. As a result of their inability to overcome obstacles, deterministic algorithms often get locked in local minima. Deterministic methods include things like energy-minimizing approaches and MD simulations [200].

### 4.4. Scoring functions

A scoring function that assesses the complex's energy affinity is used to calculate the docking score. These grading criteria may be based on consensus dock, empirical data, knowledge, or molecular mechanics [163]. The development of reliable scoring procedures and systems is of primary importance. Free-energy simulation techniques have been developed to model protein-ligand interactions quantitatively and predict affinities [202], [203]. However, evaluating several protein-ligand complexes is still impracticable due to these calculations' high cost and inconsistent accuracy. Several physical processes influence molecular recognition that is not fully accounted for by the scoring methods used in docking algorithms, which score modeled complexes utilizing a range of assumptions and simplifications [187]. In general, three distinct scoring functions exist.

### 4.4.1. Force-field-based scoring functions

Molecular mechanics is a method that approximates the handling of molecules according to the rules of classical mechanics to speed up computational times for quantum mechanical computations [204]. In molecular mechanics, the potential energy of a system with both bound (intramolecular) and unbonded (intermolecular) components is approximated by a force-field. In molecular docking, it is possible to consider the nonbonded components and ligand-bonded terms, mainly the torsional terms. The Coulomb function (which defines the electrostatic potential) and the van der Waals term (which describes the Lennard-Jones potential) can be modified by adding a distance-dependent dielectric to mimic the solvent effect. Force-field scoring functions have been changed to include new concepts, such as solvation terms [200]. GoldScore [205] and AutoDock [206] are two examples of force field-based scoring functions [196]. There is a wide variety of force-field scoring functions depending on different sets of parameters. For instance, the AMBER force field [207] is the foundation for AutoDock, while the Tripos force field [208] provides the basis for GoldScore. Even though functional shapes frequently resemble one another.

### 4.4.2. Empirical scoring functions

Initially suggested by Böhm, these scores are the product of many parameterized factors, including van der Waals, electrostatic, desolvation, hydrogen bonds, hydrophobicity, and entropy, that are fitted to replicate experimental data, such as binding energies and conformations [209]. Empirical scoring functions are constructed based on the hypothesis that binding energies may be expressed as the product of unrelated variables. The first illustration of an empirical one was the LUDI scoring function [210]. GlideScore [211], [212], ChemScore [213], and PLANTS$_{CHEMPLP}$ [214] are other practical scoring methods.

### 4.4.3. Knowledge-based scoring functions

These techniques assume a correlation between positive interactions and ligand-protein connections that are statistically more investigated. The energy element is derived from the frequency of interactions between ligand and protein atom pairs, calculated using a database of structures as a starting point. The score for a given pose is calculated by adding the sums of the energy above components for each atom pair in the ligand and protein [187], [196]. Examples of knowledge-based scoring functions include DrugScore [215], [216] and GOLD/ASP [217].

### 4.4.4. Consensus scoring functions

Another technique involves combining various scoring mechanisms to get what is known as consensus scoring [218]. Consensus scoring integrates data from many scores to counteract inaccuracies in single scores and increase the likelihood of finding 'true' ligands. X-CSCORE [219], which incorporates scoring features similar to GOLD, DOCK, ChemScore, PMF, and FlexX, is an excellent example of a consensus scoring system. Nevertheless, if terms in several scoring systems have a strong correlation, the potential utility of consensus scoring may be restricted as this could magnify computation errors rather than balance them [187], [196]. In addition, novel scoring functions, such as those based on machine learning technologies, interaction fingerprints, and quantum mechanical scores, have been established [198].

## 4.5. Quantum Mechanics/Molecular Mechanics (QM/MM)

QM/MM docking is a hybrid computational methodology that integrates the high-level accuracy of quantum mechanical (QM) approaches with the computational efficiency of molecular mechanics (MM) models. This technique is particularly critical in drug discovery, as it enables a more accurate characterization of the interactions between a ligand and its biological target. It is especially advantageous when conventional force fields are inadequate to account for key quantum phenomena, including polarization, charge transfer, and the formation or cleavage of chemical bonds during ligand binding [220], [221]. QM/MM proves exceptionally valuable in contexts demanding high precision, such as enzyme catalysis, active sites containing metal ions, or reactions involving bond formation and cleavage. These are areas where traditional MM docking or molecular dynamics simulations often fall short, primarily due to their limitations in accurately representing electronic structure changes crucial to these processes [222], [223].

In QM/MM docking, the system is divided into two distinct regions to optimize computational efficiency and accuracy: the QM region and the MM region. The QM region typically includes the most critical elements of the system, such as the ligand binding site, catalytic residues, or metal ions, and their immediate surroundings. In this region, quantum mechanical calculations accurately describe the electronic structure, often utilizing methods like density functional theory (DFT) or semi-empirical quantum approaches. These calculations capture essential electronic effects, including polarization, charge transfer, and bond formation or cleavage. Conversely, the MM region involves the larger part of the system, such as the protein backbone, solvent molecules, and more distant residues, which are less relevant to the electronic structure but essential for the system's overall dynamics. Here, classical molecular mechanics is used, applying force fields to describe non-bonded interactions and structural flexibility efficiently. The core of QM/MM docking lies in the interface between the QM and MM regions. Sophisticated algorithms ensure the smooth coupling of quantum and classical descriptions, allowing the seamless exchange of energy and forces between the two regions without introducing significant artifacts. This hybrid method offers an efficient and accurate approach for simulating large biological systems, particularly in areas where classical methods fail to capture critical electronic details. QM/MM docking strikes a balance between computational feasibility and high accuracy, especially when modeling enzyme active sites, metal-containing proteins, and ligand interactions with catalytic residues [220], [222], [224].

Several recent studies have utilized QM/MM docking to tackle complex challenges in drug discovery where traditional approaches are insufficient. For instance, QM/MM has been applied to the investigation of CYP450 inhibitors, providing more profound insights into the coordination chemistry of the heme iron involved in drug metabolism, ultimately contributing to the development of safer inhibitors [225], [226]. Furthermore, QM/MM has been instrumental in optimizing inhibitors targeting the SARS-CoV-2 main protease (Mpro), offering quantum-level insights into inhibitor binding, crucial for COVID-19 therapeutic development [227], [228]. In the context of kinase inhibitors, QM/MM facilitated the modeling of phosphoryl transfer reactions, resulting in the design of more effective drugs [229], [230], [231]. Similarly, QM/MM approaches have enhanced the design of DNA topoisomerase inhibitors by accurately simulating bond-breaking and bond-forming processes [232], [233]. Additionally, QM/MM has been applied to the study of beta-lactamase inhibitors, aiding in the design of more effective inhibitors to combat antibiotic resistance [234], [235], [236].

QM/MM docking offers distinct advantages over traditional molecular docking techniques by incorporating quantum mechanics into the most critical region of the system, such as the binding site. This allows QM/MM docking to capture essential changes in the electronic structure, including polarization and charge transfer, which are crucial for accurate predictions of ligand binding. The approach balances computational efficiency with accuracy by confining the quantum mechanical calculations to a small, relevant portion of the system, while treating the remainder with molecular mechanics. This makes QM/MM docking less computationally intensive full quantum mechanical simulations [220]. QM/MM docking is particularly effective in cases where traditional methods are inadequate, such as in enzyme catalysis, metalloenzyme interactions, or reactions involving bond formation and cleavage [222]. However, QM/MM is still more computationally demanding than purely molecular mechanics-based docking, and determining the appropriate division between the QM and MM regions can be complex, requiring specialized expertise in both areas. Additionally, the availability of QM/MM docking tools in software is limited, and those that do offer such capabilities often require advanced knowledge of quantum chemistry, which restricts its broader application [220], [237].

### 4.6. Density Functional Theory (DFT) in Drug Design

DFT is a quantum mechanical approach extensively employed in drug design to determine the electronic structure of molecules, providing critical insights into molecular interactions at the atomic scale. DFT plays a pivotal role in predicting drug candidates' reactivity, stability, electronic properties, and biological targets.

It allows researchers to model complex systems, such as enzyme-ligand interactions, with a high degree of accuracy, which is essential for understanding the binding mechanisms of drugs to their targets. DFT accounts for key quantum effects, including electron density distribution, charge transfer, and polarization, which are often beyond the capabilities of classical methods such as molecular mechanics. This makes DFT particularly valuable for studying chemical reactions, including forming and breaking covalent bonds between drugs and their target proteins and optimizing drug candidates to enhance their efficacy [194], [238].

Recent applications of DFT in drug design have led to notable progress across various therapeutic areas. In COVID-19 drug development, DFT has been employed to investigate potential SARS-CoV-2 main protease (Mpro) inhibitors, a key enzyme in viral replication. These studies have assessed small molecules' binding energies and electronic properties, aiding in identifying promising protease inhibitors and providing crucial insights into their interactions with Mpro for more precise drug design [239], [240], [241]. Furthermore, DFT has demonstrated its utility in studying antiviral drugs, especially nucleoside analogs. By analyzing these compounds' electronic properties and stability, DFT has contributed to the development of more effective and stable antiviral therapies [242], [243]. In cancer therapy, DFT has been applied to examine DNA-intercalating agents, chemotherapeutic drugs that bind to DNA. Recent research has focused on the electronic structure of drug-DNA complexes, improving the understanding of their interactions with the DNA backbone. This has enabled the optimization of these interactions, enhancing the efficacy of the drugs while reducing off-target effects [244], [245], [246]. One of the key advantages of DFT is its relatively low computational cost compared to other quantum mechanical methods while maintaining a high degree of accuracy. This balance makes DFT particularly suitable for investigating large biomolecular systems, allowing researchers to study drug-target interactions with greater depth and precision. However, DFT also has limitations. Its accuracy largely depends on the choice of exchange-correlation functional, and it can sometimes encounter difficulties with systems that involve strong correlation effects or dispersion interactions, both of which are critical in biomolecular complexes. Despite these challenges, recent advancements in DFT methods, such as developing hybrid functionals, have significantly enhanced their reliability for drug design applications [247], [248], [249], [250].

## 5. Drug-likeness

In the last 80 years, the FDA has approved roughly 1400 small-molecule oral medicines for human use. More than 40% of potential drugs never reach the market because they lack desirable biological and pharmaceutical qualities, sometimes called drug-likeness, such as solubility, sufficient chemical stability, metabolic activity, and permeability [251]. Based on the structures and features of existing drugs and drug candidates, the therapeutic likeness is commonly used in the discovery phase of drug development to exclude potentially harmful compounds [252]. Lipinski and his Pfizer co-workers conducted a seminal study evaluating the chemical characteristics of approved and experimental oral medicines and came up with the 'rule of five' (RoF) in 1997 [253]. To determine if a chemical compound contains the characteristics of a particular biological or pharmacological activity, The qualities of a medicine that allow for its oral administration in humans, as defined by Lipinski's Rule of Five, are evaluated [253], [254]. Lipinski's Rule states that a molecule is a drug if it meets the conditions shown in Figure 11.

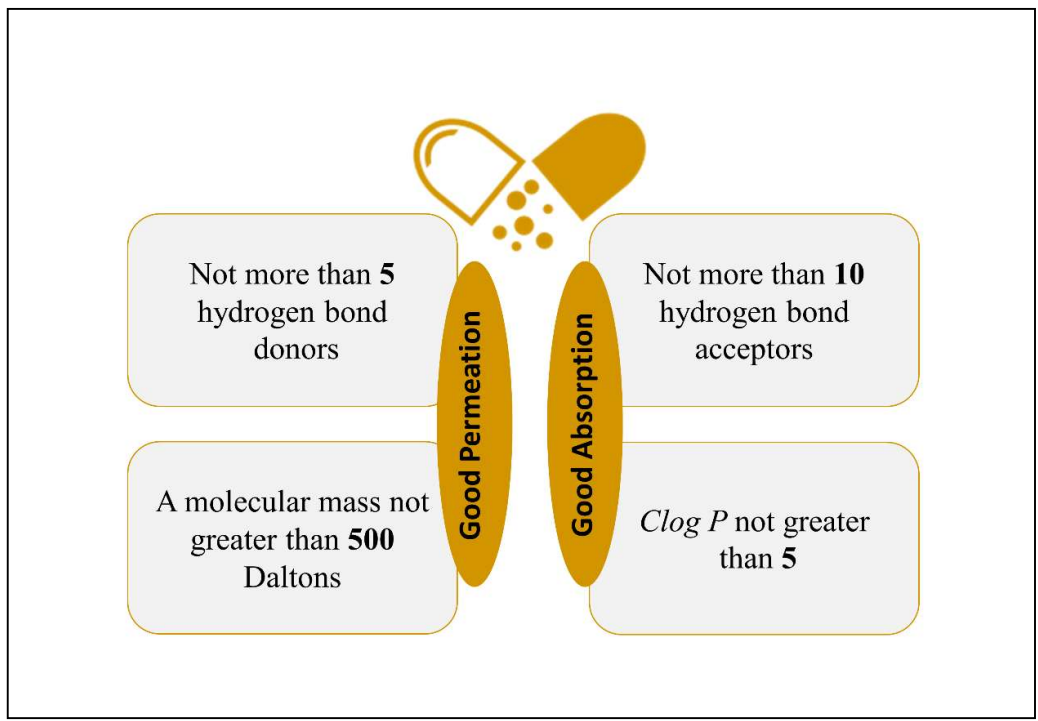

**Figure 11:** Criteria of Lipinski's Rule of Five

Analysis of successful and unsuccessful clinical candidates has further helped solidify the RoF's foundational principles. However, there is now more evidence to suggest that high lipophilicity is the primary physicochemical characteristic that causes clinical failure. That size alone is not a reliable indicator of success or failure in clinical trials [255]. Natural products and semi-synthetic chemicals continue to be an essential source of pharmaceuticals, with around 30 percent of novel chemical entities created from them in the last 40 years [256].

Natural medications must exist in a drug-like chemical space, although they do not necessarily follow the RoF. Lipinski and colleagues acknowledged this fact in their initial article. They classified these natural product deviations using a fifth criterion: the RoF does not apply to chemical classes that are biological transporter substrates. Nevertheless, rather than being more efficiently transported into cells, some of the more complex natural compounds' ability to bypass the RoF is likely due to their unique molecular design and functional group arrangement [257]. Orally active natural compounds such as paclitaxel, rapamycin, and vinorelbine (Figure 12) are potential instances of this. They have been shown to reach cells by passive diffusion despite having molecular weights more significant than 750 Dalton.

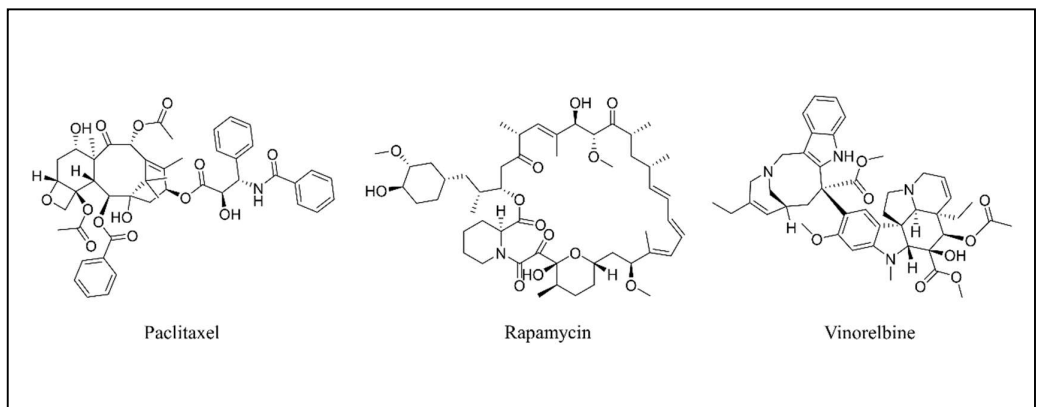

**Figure 12:** Natural origin drugs that violate the Lipinski rule of five

Predicting these features is crucial because compounds with problems relating to the pharmacokinetic parameters frequently need additional research before being approved by the national regulatory authority. Hughes [258] proposed the Pfizer 3/75 rule, which stated that drugs with clog P < 3 and polar surface area (PSA) > 75 Å exhibited decreased toxicity and, consequently, more excellent safety in pre-clinical testing. The total of polar atom surfaces in a molecule is called the polar surface area, and research has demonstrated that PSA correlates with molecular transport characteristics such as intestine absorption. The 3-dimensional shape of the molecule must be taken into account, and the surface must be estimated, calculating PSA time-consuming and computationally challenging. In 2000, Ertl and his co-worker created the TPSA method, which uses topological data to calculate PSA. TPSA offered a rapid virtual bioavailability screen of many different medications and was focused on the aggregation of surface responses of the polar components inside a molecule [259].

## 6. ADMET Properties

Pharmacology, which defines how a drug interacts with an organism, plays a significant role in discovering and developing new drugs. There are two main categories within pharmacology: pharmacodynamics (PD), which studies the effects of drugs on the body, and pharmacokinetics (PK), which investigates how the body interacts with and processes drugs [260]. The primary PK mechanism is ADME, which is then supported by toxicity (ADMET), as illustrated in Figure 13. The drug must first be sufficiently absorbed by the body and delivered to the intended location without being metabolized or eliminated. On the other hand, the objective of toxicology is to make sure there are no harmful negative effects [108], [261]. It is not guaranteed that a chemical will have the desired impact *in vivo* just because it has been shown to bind to a target of interest with specificity and to be active *in vitro*.

According to Hodgson (2001), a chemical, regardless of how specific or active its action, cannot be considered a drug unless it can enter the body in the right way (absorption), move throughout the proper areas of the body (distribution), be metabolized in a fashion that does not immediately erase its activity (metabolism), and discarded in the body safely (excretion). A drug must enter the body, move inside, hang out for a while, and then leave' [261]. Following lead discovery or lead design, much focus is placed on enhancing the compounds' *in vivo* ADMET characteristics without reducing their biological activity. Pharmaceutical companies and universities have been attracted to *in silico* ADMET property evaluation because of advances in bioactivity, property data, and machine learning methodologies for many years. Multiple ADMET-based criteria are commonly utilized early in the discovery process to decrease the number of compounds that need to be analyzed and save time and money. Thus, this essential drug discovery and design feature may be predicted using computational techniques.

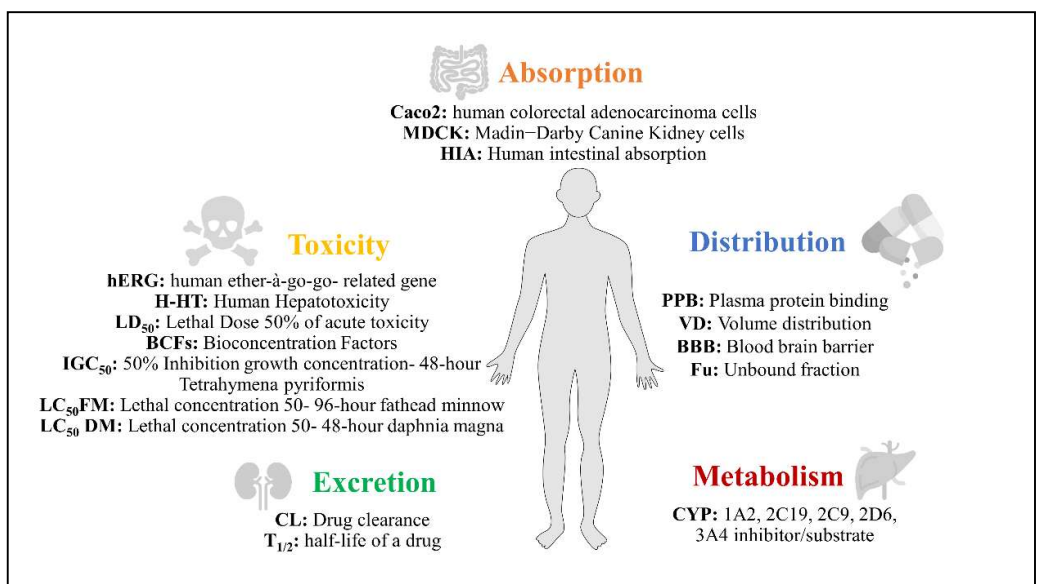

**Figure 13:** Explanation of ADMET characteristics

## 6.1. Absorption

Potential drugs face their first challenge in absorption since they must enter the bloodstream before being active within the body [262]. Drug absorption is intricately connected to numerous features: Membrane Permeability (Caco2 and MDCK) and Human Intestinal Absorption (HIA) are two representative qualities that are not only extensively researched but are also closely related to absorption [75]. Membrane permeability is a simplistic and potent physicochemical parameter for correctly predicting absorption. Two major *in vitro* permeability tests may mimic and predict drug absorption. The caco-2 permeability test and Madin-Darby canine kidney (MDCK) cell line are the first and second, respectively. Several attempts have been undertaken in recent years to evaluate the absorption capacity of drugs using machine learning and molecular descriptors [263], [264], [265], [266]. Procedures involving active transport, carrier-mediated uptake, or passive diffusion are necessary for orally administered drugs to enter the systemic circulation after crossing intestinal cell membranes. Caco-2 cell lines have been widely used to evaluate *in vivo* drug permeability because of their morphological and functional similarities to the human intestinal epithelium. As a result, Caco-2 cell permeability has also proven to be a significant indicator of a promising therapeutic molecule. The *in vitro* gold standard for determining how effectively substances is absorbed into the body is known as the apparent permeability coefficient or MDCK.

Moreover, HIA is the most crucial factor in the oral administration of drugs [267]. Additionally, it has been demonstrated that oral bioavailability and intestinal absorption are closely related, and to some extent, HIA can serve as a substitute indicator for oral bioavailability. Some prediction models for membrane permeability and human intestinal absorption have been established recently using data. The intrinsic permeability of Caco-2 cells must be measured. Fredlund *et al.* [263] created an *in vitro* assay, and they made prediction models using the data obtained from the experiment. In addition, Lanevskij *et al.* [265] proposed a nonlinear regression model that could be adjusted for the 1,366 Caco-2 cell permeability data they gathered from various sources. Ponzoni *et al.* [268] have constructed a robust machine learning model employing artificially produced and learned chemical descriptors on a dataset of 202 compounds. Wang *et al.* [269] and Yang *et al.* [270] tackled the data-imbalance issue using a tweaked algorithm and various sampling met. Molecular descriptors and feature selection were employed in both investigations; however, the latter created the method to identify the best ensemble model and feature sets.

### 6.2. Distribution

A drug must be delivered to the intended site of action after administration or absorption into the bloodstream to be effective. Distribution is the name given to this aspect of the drug. Numerous researchers make numerous attempts to predict the rate at which possible pharmaceuticals will be distributed because the distribution of a drug is a complicated function of various variables. This section discusses studies using typical distribution-related parameters, such as plasma protein binding (PPB) rate, volume distribution (VD), blood-brain barrier (BBB) permeability, and unbound percentage (Fu). The PPB rate is the proportion of a given molecule bound to plasma proteins such as human serum albumin, alpha-acid glycoprotein, and lipoprotein [75]. The rate at which drugs bind to plasma proteins is vital in predicting distribution since the drug molecules have no pharmacological effect when they form the protein-ligand complex, even after reaching the target tissues [271]. Using *in vivo* or *in vitro* observed data, scientists are still actively researching *in silico* PPB prediction models [272], [273], [274], [275], [276]. Wang *et al*. [273] assembled an extensive PPB dataset from various sources, including the DrugBank database. They built a prediction model using an ensemble of machine learning models and an optimized feature set. Three data sets from scholarly research and open-access databases were utilized by Sun *et al.* [274] to develop a robust prediction model. Several machine learning techniques and synthetic molecular descriptors are used in the modeling process. Toma *et al.* [275] obtained *in vivo* data for PPB prediction models. To train the Random Forest (RF) model, they first created molecular 2D descriptors and features based on SMILES. Some ADME characteristics, such as the PPB rate half-life, can be predicted using a multitask DNN architecture, as proposed by Zhuyifan *et al.* [276]. The model was trained using DeepChem's molecular property benchmark dataset to make accurate predictions across an ample chemical space. However, VD is defined by comparing the drug's total body mass to its plasma concentration. It is a conceptual bridge between the prescribed dose and the first concentration found in the blood. It is a crucial metric to characterize the *in vivo* distribution of medications. Practically speaking, we can make assumptions about the distribution characteristics of an unknown chemical based on its VD value, such as its affinity for plasma proteins, its dispersion in bodily fluids, and its uptake in tissues. The blood-brain barrier (BBB) prevents harmful compounds from entering the brain via the bloodstream, making BBB permeability a significant challenge in developing CNS-targeted (central nervous system-targeted) medications. Many researchers are trying to create accurate forecasting models and spot underlying patterns. There are machine learning-based [277–279] and deep learning-based [277], [278], [279], [280], [281] methods for BBB prediction. Toropov *et al.* [278] built their model using SMILES-based characteristics extracted from the CORAL database. To adapt their model to a rather extensive dataset encompassing 2,358 and 3,538 chemicals, Wang *et al.* [279] and Yuan *et al.* [280] instead relied on more conventional 2D molecular descriptors and fingerprints. Miao *et al*. [281] have presented a fascinating study on BBB prediction using deep learning. The author developed a new drug-phenotype feature by merging data from the SIDER database and the Medical Dictionary for Regulatory Activities. Furthermore, Fu implies that most drugs in plasma will maintain a stable equilibrium between the unbound form and the bound form of serum proteins. How well a medication disperses throughout the body and can pass cellular membranes may be affected by how well it binds to blood proteins [75].

### 6.3. Metabolism

Metabolic enzymes play different roles in the biotransformation process known as metabolism. The drug may alter the metabolic pathway that regulates the activation or excretion of other pharmaceuticals, or it may change into new chemicals that can be activated or excreted [271]. Phase I (oxidative reactions) and Phase II (conjugative reactions) are two general categories that can be used to categorize drug metabolism reactions based on the chemical basis of biotransformation. The human Cytochrome P450 (CP450) family (phase I enzymes) metabolizes around two-thirds of all known medicines, with the five isozymes 1A2, 3A4, 2C9, 2C19, and 2D6 responsible for over 80% of this process. These CYPs, accountable for phase I reactions, are concentrated primarily on the liver. The primary focus of drug metabolism studies is predicting the CYP450 enzyme's role in drug metabolism. Predictions for CYP450 substrates and inhibitors are the two main streams of CYP450-related prediction. Because the CYP450 substrate can impact pharmacological efficacy, excretion, and toxicity, it is crucial to predict it.

Similarly, CYP450 inhibitor prediction is essential since it directly impacts drug-drug interactions and the resulting toxicities. Several research teams have recently released CYP450 substrate prediction studies [282], [283], [284]. Hunt *et al.* [282] proposed using a multiclass prediction model to determine which CYP450 isoform metabolizes the query chemical. The team used common chemical descriptors and information on 633 compound isoform pairings to construct the multiclass RF model. The 'CypReact' prediction tool was developed by Tian *et al.* [283] with the help of 1,632 gathered compounds, including 679 CYP450 reactants. A unique cost-sensitive meta-learning approach, the Learning Base Model, was used to determine the ideal feature sets and classifiers for any budget. They made use of a variety of physical and chemical fingerprints as a criterion. Shan *et al.* [284] suggested a multi-label classification model using 1,299 compound isoform data pairs.

### 6.4. Excretion

Excretion refers to the process of eliminating undesirable material from the body. This step is important since the drug's dose is based on its excretion factor. Drug elimination occurs complexly depending on some chemical and physiological factors and the drug metabolism process. As a result, only a few recent studies have been able to predict the features related to excretion accurately. This section covers research on drug clearance (CL) and drug half-life ($t_{1/2}$). The clearance of a drug is a significant pharmacokinetic parameter that, along with the volume of distribution, determines the half-life and, thus, the dosage frequency. While the half-life of a medication is a combination concept involving clearance and volume of distribution, it may be preferable to have accurate estimations of these two qualities instead. A study to predict medication plasma clearance was created by Zhivkova *et al*. [285]. A linear regression model with well-selected characteristics was built using a genetic algorithm and data representative of 659 medicines. Wakayama *et al.* [286] proposed a prediction model that assesses major drug clearance routes. Two hundred forty-nine medications were used, and data from nine major clearance routes (such as renal, OATP, and CYP450-related) was included.

### 6.5. Toxicity

Unwanted adverse effects, including drug toxicity, may result in enormous expenditures if not adequately investigated during drug development. The late-stage failure rate can be reduced by using *in silico* toxicity prediction, which has been studied extensively since drug toxicity is the most crucial aspect of drug development. Although many investigations into toxicity prediction have been conducted, those focusing on the prediction of human ether-à-go-go related gene (hERG)-associated cardiotoxicity are of particular interest because many currently available medications are removed from the market because of this toxicity. Therefore, the hERG K+ channel is a vital antigen to consider early in drug development. Recently, it has been the subject of intensive investigation by scientists hoping to use it to anticipate hERG-related toxicity using computer simulations [287], [288], [289], [290], [291], [292], [293]. To verify their model, Siramshetty *et al.* [288] pre-processed 5,804 chemicals from the ChEMBL database and gathered information from the scientific literature. Machine learning methods and several molecular fingerprints were employed. The authors compared models trained with various cut-offs since hERG blockers lack a common cut-off potency. The hERG blocker categorization model was proposed by Ogura *et al*. [292] using the hERG-integrated database from their earlier work [294].

## 7. Molecular Dynamics Simulations

Molecular Dynamics (MD) simulations are a crucial computational tool for studying the physical foundations of biological macromolecule structure and function. MD simulation is an essential tool for theoretical studies of biological molecules. Since molecular systems often contain particles, analyzing such a complex system is impossible. Using numerical methods, molecular dynamics simulation can avoid such analytically challenging issues. With MD, it is possible to simulate how molecular systems behave dynamically over time. MD simulations properly treat water and intrinsic receptor flexibility, representing the dynamic character of ligand-receptor interactions in contrast to a static description of ligand-receptor interactions. MD is an excellent method for studying the atomic-level interactions fundamental to ligand affinity and selectivity toward its intended destination (target) [196].

In most cases, an MD simulation's underlying algorithm is made by

1. identification of the starting velocities and locations of each atom,
2. use of inter-atomic potentials to calculate the force applied on the analyzed atom,
3. development of atomic locations and velocities during a limited period.

In step 2, the updated location and velocity are utilized as inputs, and in step 3, a new time step is generated [295]. Molecular dynamics is often used to study biological molecules and their complexes from a structural, behavioral, and thermodynamic perspective. It provides in-depth information about shifts and changes in protein structure. This approach can also be used to explore the structural information of nucleic acids. Additionally, the effects of general protein structural modifications can be studied about solvent molecules [296], [297].

### 7.1. Steps Involved in Constructing the Model System for MD Simulation

A well-prepared model system must be constructed following several key steps to perform an MD simulation. First, the process begins with structure preparation, where a high-resolution 3D structure of the molecule is obtained, typically from the Protein Data Bank. In cases where the structure is incomplete or missing segments, homology modeling tools are employed to fill in missing parts, and all atoms, including those in loops or residues, are added using molecular modeling software (as described in Section 3.1.6. Protein Structure Prediction). After the structure is prepared, the system undergoes system solvation, in which the biomolecule is placed in a solvent box, typically filled with water, to replicate its natural environment. Water models such as TIP$_3$P or SPC are used for the solvent environment, and the size of the solvent box is carefully chosen to avoid interactions between periodic images of the system. Next, ion addition and neutralization are performed to simulate physiological conditions. Counterions like Na$^+$ and Cl$^-$ are added to neutralize the system's net charge, while additional ions may be introduced to mimic the ionic concentration of biological environments, ensuring realistic simulations. Following this, an appropriate force field (such as AMBER [298], CHARMM [299], or GROMOS [300]) is assigned to define the interactions between atoms in the system. Force fields handle parameters such as bond lengths, angles, torsions, and non-bonded interactions, thus providing a physical basis for the system's behavior (see Section 4.4.1. Force-field-based scoring functions). Once the force field is applied, the system undergoes energy minimization to remove unfavorable steric clashes and high-energy conformations that may have arisen during model construction, ensuring that the system starts in a stable, low-energy state. After energy minimization, the equilibration phase is initiated. During equilibration, the system is gradually brought to the desired temperature and pressure, typically using the NVT (constant number of particles, volume, and temperature) or NPT (constant number of particles, pressure, and temperature) ensembles. This phase ensures that the system achieves stability before the production run. Finally, the production MD simulation begins after equilibration. In this phase, the atomic motions of the system are simulated over time, typically ranging from nanoseconds to microseconds. During the production phase, the system's dynamic properties, including ligand-receptor interactions, structural flexibility, and energy fluctuations, are observed and analyzed, providing critical insights into molecular behavior [104], [106], [301], [302], [303], [304], [305], [306].

## 7.2. Membrane Protein Simulation Process

Simulating membrane proteins requires additional considerations compared to standard biomolecular simulations due to the intricate nature of the lipid environment surrounding these proteins. The first step involves inserting the membrane protein into a lipid bilayer, often using a pre-constructed bilayer such as POPC or DPPC, miming the biological membrane. Tools like CHARMM-GUI [307] or MemProtMD [308] are commonly utilized to accurately position the membrane protein within the lipid bilayer. Subsequently, lipid-specific force fields, such as CHARMM36 [309] or Lipid14 [310], are employed to model the interactions between the protein and the lipid molecules, ensuring that hydrophobic and hydrophilic interactions are accurately captured within the membrane environment. Once the protein is embedded, solvation and ion addition are conducted to solvate the external and internal environments of the membrane protein, such as any pores it may have. Ions are introduced to mimic physiological conditions, as is done in typical MD simulations. After solvation, the system undergoes energy minimization and equilibration to resolve any unfavorable interactions between the lipids and the protein. This equilibration process is typically performed in stages, allowing the lipid bilayer and protein to gradually adjust under controlled conditions, thereby ensuring a realistic membrane environment. Finally, the production MD simulation is performed under the NPT ensemble, maintaining constant pressure to preserve the surface tension and thickness of the lipid bilayer. Throughout the simulation, the interactions between the membrane protein and the lipid environment are closely monitored, providing essential insights into the protein's function and dynamic behavior over the simulation period [311], [312], [313].

## 7.3. Data Analysis in MD Simulations

Figure 14 shows the basic steps involved in running an MD simulation. It begins with building a model system that restores missing segments, identifies protonation states, neutralizes the system by adding counter ions, chooses the simulation box's size, and solvates the solutes with explicit solvent molecules. The system is then equilibrated and energy minimized until its properties no longer vary with time. After the stabilization of the system, a production run is conducted for a sufficient amount of time to generate trajectories, which are then examined to determine their relevant features. Typical MD simulation data analysis needs in-depth familiarity with the molecular system under study and a high level of technical skill. The researcher aimed to incorporate useful analysis tools to analyze protein-ligand MD trajectories better.

Common metrics for analyzing MD simulation data include Root Mean Square Deviation (RMSD) and Root Mean Square Fluctuations (RMSF). The RMSD has already been precomputed based on atom selections made in advance. It is possible to gain insight into the protein's structural conformation during the simulation by monitoring the RMSD to see if the protein is stable and if the simulation has reached equilibrium. The ligand's relative stability (concerning the protein) and its internal conformational development can be inferred from its RMSD. Furthermore, the RMSF of protein residues allows for observing regions along the protein that fluctuated more during the simulation. Moreover, the Simulation Interactions Diagram (SID) highlights the protein residues interacting with the ligand. This tool allows you to observe how fluctuations correspond to secondary structural elements in your simulation by superimposing alpha-helix and beta-strand sections of the protein onto the plot, as these regions are more rigid than the unstructured loop region(s).

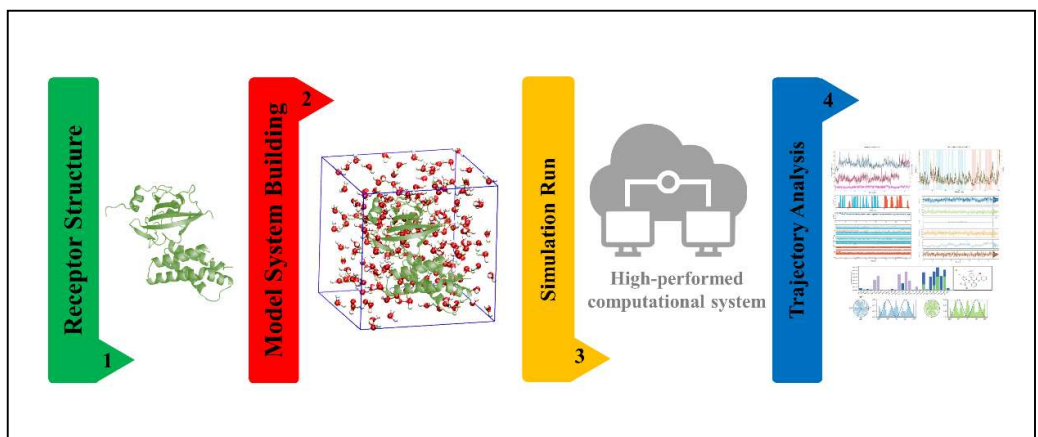

**Figure 14:** Basic steps involved in the MD simulation process

## 8. Binding Free Energy Calculation

Molecular dynamics-based binding free energy estimates can improve the accuracy of ranking the hit compounds, significantly affecting the hit identification phase. Many essential biological activities rely on molecular recognition [189]. The free energy of binding ($\Delta G_{bind}$) between two interacting molecules in a cell measures the importance of these interactions, which trigger various biological processes. Change in free energy ($\Delta G$) is a metric that can characterize a system's thermodynamic and kinetic features. (In *in silico*, the process of a ligand interacting with a protein in a solvent). '$\Delta G$' refers to the quantity of energy given off or taken in during a chemical reaction. Suppose the chemical process (In *in silico*, the interaction of a ligand to its receptor) is spontaneous. In that case, the ultimate $\Delta G$ value at the state of equilibrium will be negative, but if the method is not spontaneous, the value will be positive. Typically, the Gibbs function change is used to calculate $\Delta G$:

$$\Delta G = \Delta H - T\Delta S \quad \ldots\ldots\ldots 1$$

$\Delta H$ is the enthalpy change, T is the Kelvin temperature, and $\Delta S$ is the entropy change.

In addition, the stability of a system (In *in silico*, receptor`s ligand affinity) can be measured by the system's entropy, or $\Delta G$ [314]. To calculate the binding constant $K_b$ from the conventional binding free energy $\Delta G°_{bind}$, the following formula is used:

$$\Delta G°_{bind} = -RT\ln K_b \quad \ldots\ldots\ldots 2$$

The usual parameters for measuring $\Delta G°$ are 1 atm pressure, room temperature (298.15 K), and concentrations of 1 M for both the protein and ligand. The universal gas constant, R, has a unit of cal K$^{-1}$ mol$^{-1}$. Compared to $K_d$, the dissociation constant, $K_b$ is a ratio of the association and dissociation kinetic rate constants ($k_{on}$ and $k_{off}$). Since free energy may be regarded as a state function, the ligand's binding affinity ($K_d$) can be determined by finding the difference between the energies of the initial (unbound) and final (bound) states in a thermodynamic equilibrium [314]. This is true even if the precise mechanism leading from the initial to the final state is unknown. Extensive MD simulations are required to obtain the unbound and the bound equilibrium states, which is necessary for a computationally realistic estimation of $\Delta G_{bind}$. Evaluating $\Delta G_{bind}$ for a series of ligands against a particular target protein is a common way to identify compounds with the highest binding affinities, which can lead to the development of more potent drugs. Therefore, when the drug design and docking-based virtual screening procedures are completed, $\Delta G_{bind}$ calculations, for which several computational techniques have been created, are often undertaken [314].

The long-standing Molecular Mechanics Poisson-Boltzmann Surface Area (MM-PBSA) and MM-Generalized Born Surface Area (MM-GBSA) methods are examples of computationally less expensive end-point methods [315]. The docked poses of the most promising ligands are re-ranked using post-docking analysis (using MM-PB(GB)SA techniques) for accounting for the binding free energies of the compounds that were hit in a computational screen [316], [317]. The MM-PB(GB)SA approach to estimating the binding free energy solely uses the system's unbound and bound states (extremes). The free energy of a ligand binding to a protein can be determined using the following formula [318]:

$$\Delta G_{bind} = G_{complex} - G_{protein} - G_{ligand} \quad \ldots\ldots\ldots 3$$

Each ligand, protein, and complex's free energy (G) is derived from the sum of its gas-phase molecular mechanics energy ($E_{MM}$) (which includes internal van der Waals and electrostatic energies, as well as bound and non-bonded electrostatic energies) and its solvation free energy ($G_{solv}$), which includes polar and non-polar contributions [319]. Furthermore, at a given temperature T, the entropy (S; includes effects from translation, rotation, and vibration):

$$G = E_{MM} + G_{solv} - TS \quad \ldots\ldots\ldots 4$$

Two common schemes, known as the 'three-trajectory scheme' and the 'single-trajectory scheme', have long been used to determine the MM-PB(GB)SA. The former is relatively accurate but computationally expensive because it uses snapshots from three MD trajectories: apoprotein, free ligand, and the ligand-protein complex. However, a single trajectory approach can significantly reduce computational time by requiring only a single MD simulation of the ligand-protein combination. However, the latter method ignores the possibility that the protein and ligand may relax their structures following interaction. The MM-PB(GB)SA calculation also depends on the time spent in simulation, the solute dielectric constant, the force field used, the net charge of the system, and the solvent model, in addition to the strategy employed. Multiple separate short MD simulations have been observed to yield more accurate estimates of the $\Delta G_{bind}$ constant than a single lengthy MD trajectory. It is computationally difficult to take enough MD snapshots to get a good approximation of absolute ΔG$_{bind}$. As a result, in structure-based drug discovery [320], relative ΔG$_{bind}$ is usually all that's needed to rank compounds in terms of their affinity for the target protein, and it provides a more accurate forecast than the absolute ΔG$_{bind}$ calculations. Numerous studies have used the MM-PB(GB)SA method, which has been used to create anticancer compounds, antiviral, antibacterial, antipsychotics, and antiparasitic drugs [321].

## 9. Conclusion

We examined artificial intelligence's involvement in drug development in this work. The precise stages of drug discovery and design were then described in depth. We also discussed the molecular docking procedure, which involves examining the compatibility of various molecule architectures. Following an explanation of the drug-likeness process, the ADMET features were discussed. The binding free energy was then computed. The main goal of this work was to review *in silico* drug discovery methods with an emphasis on identifying therapeutic targets where genes or proteins are connected to certain diseases. This systematic review provided a thorough assessment of A-to-Z *in silico* drug discovery strategies to systematically define the targets of bioactive compounds. The process of finding new drugs and developing them is time-consuming and expensive. It begins with target identification, follows with target validation, and then names medication candidates. Any brand-new medicine must pass rigorous preclinical and clinical testing and receive FDA clearance before it can be sold. Because experimental procedures cannot be used widely due to throughput, accuracy, and expense limitations, drug development has recently switched to *in silico* methods, including homology modeling, protein-ligand interactions, microarray analysis, and vHTS. The development of quick and precise target identification and prediction methods for the discovery relied heavily on in silico approaches. We have noticed that The drug development process poses a significant challenge for the pharmaceutical industry, as it is both time-intensive and requires the identification of new therapeutic possibilities to treat a wide range of diseases. For future reading, the authors advise the reader could optionally read the following research works wing research works [322], [323], [324], [325], [326], [327], [328], [329], [330], [331], [332], [333], [334], [335], [336], [337], [338], [339].

**Acknowledgments**

The authors thank Charmo University for their ongoing support and provision of facilities that made this research possible.

**Funding:** Not applicable.

**Conflict of interest:** None.

**Ethical approval:** This article does not contain any study with human participants or animals performed by any authors.

**Availability of data and material:** All data generated during this study are included in supplementary information files.

**Code availability:** Not applicable

**Authors' contributions:**

Conceptualisation: HOR, BAH

Data curation: HOR, AK

Software: BKA, TAR

Resources: BKA, DDG

Formal Analysis: BAH, AK

Methodology: BAH, AK

Visualisation and Validation: BAH, YAR, DDG

Writing original draft: HOR, AK

Writing review & editing: HOR, DDG, TAR